\documentclass[aps,prb,preprint,showpacs]{revtex4}

\usepackage{times}
\usepackage{graphicx,units}
\usepackage{amssymb,stmaryrd}
\usepackage{pifont,wasysym}
\newcommand{\Diamondblack}{\ding{117}}
\newcommand{\medbullet}{\ding{108}}
\newcommand{\medcirc}{\ocircle}
\usepackage[usenames]{color}

\begin{document}

\title{Tailoring electronic and optical properties of TiO$_{\text{2}}$: nanostructuring, doping and molecular-oxide interactions}    

\author{L. Chiodo$^\dagger$}
\author{J. M. Garc{\'{\i}}a-Lastra$^\dagger$}
\author{D. J. Mowbray$^\dagger$}
\author{A. Iacomino$^\ddagger$}
\author{A. Rubio$^{\dagger,\S}$} 
 \affiliation{
$^\dagger$Nano-Bio Spectroscopy group and ETSF Scientific Development
  Centre, Dpto. F{\'{\i}}sica de Materiales,  
Universidad del Pa{\'{\i}}s Vasco, Centro de F{\'{\i}}sica de
Materiales CSIC-UPV/EHU- MPC and DIPC, Av. Tolosa 72, E-20018 San
Sebasti{\'{a}}n, Spain\\
$^\ddagger$Dipartimento di Fisica ``E. Amaldi'', Universit{\`{a}} degli Studi Roma Tre,
Via della Vasca Navale 84, I-00146 Roma, Italy\\
$^\S$Friz-Haber-Institut der Max-Planck-Gesellschaft, Berlin, Germany}

\begin{abstract}
Titanium dioxide is one of the most widely investigated oxides.  This
is due to its broad range of applications, from catalysis to
photocatalysis to photovoltaics. Despite this large interest, many of
its bulk properties have been sparsely investigated using either
experimental techniques or \emph{ab initio} theory. Further, some of
TiO$_{\text{2}}$'s most important properties, such as its electronic
band gap, the localized character of excitons, and the localized
nature of states induced by oxygen vacancies, are still under
debate. We present a unified description of the properties of rutile
and anatase phases, obtained from \emph{ab initio} state of the art
methods, ranging from density functional theory (DFT) to many body
perturbation theory  (MBPT) derived techniques. In so doing, we show
how advanced computational techniques can be used to quantitatively
describe the structural, electronic, and optical properties of
TiO$_{\text{2}}$ nanostructures, an area of fundamental importance in
applied research. Indeed, we address one of the main challenges to
TiO$_{\text{2}}$-photocatalysis, namely band gap narrowing, by showing
how to combine nanostructural changes with doping. With this aim we
compare TiO$_{\text{2}}$'s electronic properties for 0D clusters, 1D
nanorods, 2D layers, and 3D bulks using different approximations
within DFT and MBPT calculations. While quantum  confinement effects
lead to a widening of the energy gap, it has been shown that
substitutional doping with boron or nitrogen gives rise to
(meta-)stable structures and the introduction of dopant and mid-gap
states which effectively reduce the band gap. Finally, we report how
\emph{ab initio} methods can be applied to understand the important
role of TiO$_{\text{2}}$ as electron-acceptor in dye-sensitized solar
cells.  This task is made more difficult by the hybrid organic-oxide
structure of the involved systems. 
\end{abstract}

\maketitle

\section{Introduction}

TiO$_{\text{2}}$ has been one of the most studied oxides over the past
few years.  This is due to the broad range of applications it offers
in strategic fields of scientific, technological, environmental, and commercial
relevance. In particular, TiO$_{\text{2}}$ surfaces and nanocrystals
provide a rich variety of suitably tunable properties from structure to
opto-electronics. Special attention has been paid to
TiO$_{\text{2}}$'s optical properties.  This is because
TiO$_{\text{2}}$ is regarded as one of the best candidate materials to
efficiently produce hydrogen via photocatalysis~\cite{Gratzel2001,
  FirstPhotocatalysis}. TiO$_{\text{2}}$ nanostructures are also
widely used in dye-sensitized solar cells, one of the most promising
applications in the field of hybrid solar cells~\cite{Gratzel2001}. 

Since the first experimental formation of hydrogen by photocatalysis
in the early 1980s,~\cite{FirstPhotocatalysis} TiO$_{\text{2}}$ has
been the catalyst of choice. Reasons for this include the position of
TiO$_{\text{2}}$'s conduction band above the energy of hydrogen
formation, the relatively long lifetime of excited electrons which
allows them to reach the surface from the bulk, TiO$_{\text{2}}$'s
high corrosion resistance compared to other metal oxides, and its
relatively low cost~\cite{Gratzel2001,
  Hoffmann1995,C-TiO2doping}. However, the large optical band gap of bulk
TiO${}_{\text{2}}$ ($\approx$ 3~eV) means that only high energy UV
light may excite its electrons.  This effectively blocks most of the
photons which pierce the atmosphere, typically in the visible range,
from participating in any bulk TiO${}_{\text{2}}$ based photocatalytic
reaction.  On the other hand, the difference in energy between excited
electrons and holes, i.e. the band gap, must be large enough
($\gtrsim$ 1.23~eV) to dissociate water into hydrogen and oxygen.  For
these reasons it is of great interest to adjust the band gap
$\varepsilon_{gap}$ of TiO${}_{\text{2}}$ into the range 1.23
$\lesssim \varepsilon_{gap} \lesssim$ 2.5~eV, while maintaining the
useful properties mentioned above~\cite{TiO2PRL}. 

With this aim, much research has been done on the influence of 
nanostructure~\cite{Structures,ExpTiO2clusters,TiO2NTs1,TiO2NTs2,Delaminated,TiO2}
and dopants
~\cite{TiO2,JACS-suil,N-TiO2NTarraydoping,N-TiO2doping,N-TiO2NTdoping,N-TiO2NTdoping2,Nambu,Graciani,TiO2PRL,zhu:226401}
on TiO${}_{\text{2}}$ photocatalytic activity.  For low dimensional nanostructured
materials, electrons and holes have to travel shorter distances to
reach the surface, allowing for a shorter quasi-particle
lifetime. However, due to quantum confinement effects, lower
dimensional TiO${}_{{{2}}}$ nanostructures tend to have \emph{larger}
band gaps~\cite{TiO2Rev1}.  On the other hand, although doping may
introduce mid-gap states, recent experimental studies have shown that
boron and nitrogen doping of bulk TiO${}_{\text{2}}$ yields band gaps
\emph{smaller} than the threshold for water
splitting~\cite{JACS-suil,N-TiO2NTarraydoping}. This suggests that low
dimensional structures with band gaps larger than about \(3.0\)~eV may
be a better starting point for doping. 

Recently, several promising new candidate structures have been
proposed~\cite{TiO2}.  These   small ($R \lesssim$ 5 \AA)
TiO${}_{\text{2}}$ nanotubes, with a hexagonal ABC PtO${}_{\text{2}}$
structure (HexABC), were found to be surprisingly stable, even in the
boron and nitrogen doped forms.   This stability may be attributed to
their structural similarity to bulk rutile TiO${}_{2}$, with the
smallest nanotube having the same structure as a rutile nanorod. 

A further difficulty for any  photocatalytic system is controlling how
electrons and holes travel through the system ~\cite{Turner}. For this
reason, methods for reliably producing both $n$-type and $p$-type
TiO${}_{\text{2}}$ semiconducting materials are highly desirable. So
far, doped TiO${}_{\text{2}}$ tends to yield only $n$-type
semiconductors.  However, it has recently been proposed that $p$-type
TiO$_{\text{2}}$ semiconducting materials may be obtained by nitrogen
doping surface sites of low dimensional materials~\cite{TiO2}. 
In this chapter we will discuss in detail the
effects of quantum confinement and doping on the optical properties of
TiO$_{\text{2}}$.  

TiO$_{\text{2}}$ nanostructures are also one of the main components of hybrid solar
cells. In a typical Gr{\"{a}}tzel cell~\cite{Gratzel2001},
TiO$_{\text{2}}$ nanoparticles with average diameters around 20 nm
collect the photoelectron transferred from a dye molecule adsorbed on
the surface~\cite{Nazeeruddin1,Nazeeruddin2}.  Such processes are
favoured by a proper energy level alignment between solid and organic
materials, although the dynamic part of the process also plays an
important role in the charge transfer.  Clearly, TiO$_{\text{2}}$'s
characteristics of long quasi-particle lifetimes, high corrosion
resistance, and relatively low cost, must be balanced with control of
its energy level alignment with molecular states, and a fast electron
injection at the interface.  

Despite all the engineering efforts, the main scientific goal remains
to optimize the efficiency of solar energy conversion into readily
available electricity.  Different research approaches have been
devoted to benefit from quantum size properties emerging at the
nanoscale~\cite{Mao,Meyer}, find an optimal donor-acceptor
complex~\cite{Mao,Meyer},  mix nanoparticles and one-dimensional
structures, such as nanotubes or nanowires~\cite{Jih-Jen,Kang}, and
control the geometry of the TiO$_{\text{2}}$
nano-assembly~\cite{Adachi}. 

A clear theoretical understanding of TiO$_{\text{2}}$'s optoelectronic
properties is necessary to help unravel many fundamental questions
concerning the experimental results.  In particular, the properties of
excitons, photo-injected electrons, and surface configuration in
TiO$_{\text{2}}$ nanomaterials may play a critical role in determining
their overall behaviour in solar cells. For TiO$_{\text{2}}$ at both
the nanoscale and macroscale regime, it is necessary  to first have a
complete picture of the optical properties in order to clarify the
contribution of excitons.  

\begin{figure}
\centering
\includegraphics[width=0.8\textwidth]{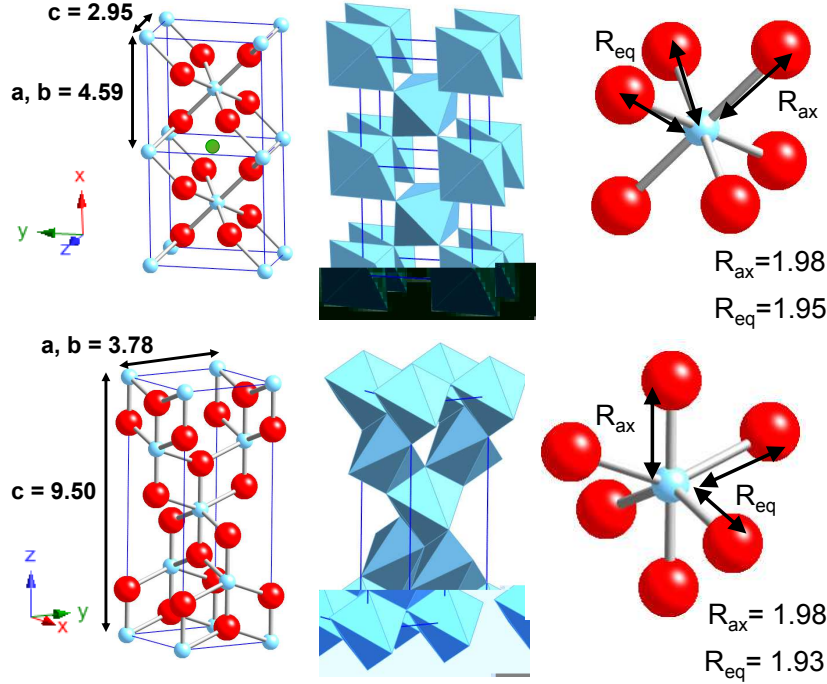}
\caption{Unit cell (left), crystallographic structure (center) and
  TiO$_6$ octahedrons (right) of rutile (top) and anatase (bottom). 
The lattice parameters in \AA~are denoted \(a\), \(b\), and \(c\), while $R_ax$ and $R_eq$ are 
the distances in \AA~between a Ti$^{\text{4+}}$ ion and its nearest and next-nearest neighbour O$^{\text{2-}}$ ions, 
respectively.   In the case of rutile the interstitial Ti impurity site is shown with a
  green circle (see left top). }\label{LFig1} 
\end{figure}

Despite the clear importance of its surfaces and nanostructures,
investigations of TiO$_{\text{2}}$ bulk (see \ref{LFig1}) electronic
and optical properties have not provided, so far, a comprehensive
description of the material. Important characteristics,such as the
electronic band gap, are still undetermined.  Most of the experimental
and computational work has been focused on synthesis and analysis of
systems with reduced dimensionality.  

The experimental synthesis and characterization of nanostructured
materials is in general a costly and difficult task. However,
predictions of a dye or nanostructure's properties from simulations
can prove a great boon to experimentalists. Modern large-scale
electronic structure calculations have become important tools by
providing realistic descriptions and predictions of structure and
electronic properties for systems of technological interest.

Although it will not be treated in this chapter, it is important to
mention the problem of electron localization in reduced
titania~\cite{GandugliaPirovano2007219}.  This will provide a glimpse
of the complexity faced, from the theoretical point of view, when
studying transition metal atoms. The localization of $d$ electrons
makes the accurate description of their exchange--correlation interaction a  
difficult task~\cite{gatti2007}.  The electron localization in
defective titania has been an open question from both experimental and
theoretical points of view, and caused much controversy during the
past few years~\cite{kruger_prl,DiValentin,minato_jcp}.  Oxygen
vacancies are quite common in TiO$_{\text{2}}$, and their presence and
behaviour can significantly affect the properties of
nanostructures. When an oxygen vacancy is created in TiO$_{\text{2}}$,
i.e.~when TiO$_{\text{2}}$ is reduced, the two electrons coming from
the removed O$^{\text{2-}}$ ion must be redistributed within the
structure.  One possibility is that these two extra electrons remain
localized onto two Ti ions close to the O$^{\text{2-}}$ vacancy. In
this way a pair of Ti$^{\text{4+}}$ ions become Ti$^{\text{3+}}$
ions. Another option is for the two extra electrons to delocalize
along the whole structure, i.e.~they do not localize on any particular
Ti ion. Finally, an intermediate situation, with one electron
localized and the other spread, is also possible.  Concerning the
TiO$_{\text{2}}$ bulk, conventional density functional theory (DFT)
calculations using either local density approximations (LDA) or
generalized gradient approximations (GGA) for the exchange-correlation
(xc)-functionals show a scenario with both electrons delocalized.  On
the other hand, hybrid functionals and Hartree-Fock calculations give
rise to a situation with both electrons localized.  For GGA+U
calculations, the results are very sensitive to the value of the U
parameter. For certain values of U both electrons remain localized,
while for others there is an intermediate
situation~\cite{finazzi_jcp,pacchioni_alone}.  

Experimentally, there are electron paramagnetic resonance (EPR) measurements suggesting that the extra
electrons are mainly localized on interstitial Ti$^{\text{3+}}$
ions~\cite{Khomenko}. These interstitial Ti$^{\text{3+}}$ ions are
impurities placed at the natural interstices of the rutile lattice
(see \ref{LFig1}) and, similarly to the Ti$^{\text{4+}}$ ions of the
pure lattice, they also form TiO$_6$ octahedrons. Recent STM and PES
experiments have shown that the interstitial Ti$^{\text{3+}}$ ions
play a key role in the localization of the electrons when a bridge
oxygen is removed from the TiO$_{\text{2}}$ (110)
surface~\cite{StefanWendt06272008}. These experiments concluded the
controversial discussion about the localization of electrons in the
bridge oxygen defective TiO$_{\text{2}}$ (110) surface (see
Refs.~\cite{kruger_prl,DiValentin,minato_jcp} for more
details). However, the problem remains unresolved for the bulk case. 

In summary, in this chapter we first analyze the full \emph{ab initio}
treatment of electronic and optical properties in \ref{DFT} and
\ref{MBPT}, before applying it to the two most stable bulk phases, rutile and anatase
in \ref{bulk}. These are also the phases most easily found
when nanostructures are synthesized. We will focus on their optical
properties and excitonic behaviour.  We then explore the possibility
of tuning the oxide band gap using quantum confinement effects and
dimensionality, by analyzing atomic clusters, nanowires and nanotubes
in \ref{nanostructure}. A further component whose effect has to be
evaluated is that of doping, which may further tune the optical
behaviour by introducing electronic states in the gap, as presented in
\ref{doping}.Combining the effects of quantum confinement and doping
is hoped to produce a refined properties control. In \ref{cells}, we
report some details on modeling for dye-sensitized solar cells, before
providing a summary and our concluding remarks. 

\section{Ground state properties through Density Functional
  Theory}\label{DFT}

DFT is a many-body approach, successfully used for many years to
calculate the ground-state electronic properties of many-electron
systems. However, DFT is by definition a ground state theory, and is
not directly applicable to the study of excited states. To describe
these types of physical phenomena it becomes necessary to include
many-body effects not contained in DFT through Green's function
theory. The use of many-body perturbation theory~\cite{GW}, with DFT
calculations as a zero order approximation, is an approach widely used
to obtain quasi-particle excitation energies and dielectric response
in an increasing number of systems, from bulk materials to surfaces
and nanostructures. We present a short, general discussion of the
theoretical framework, referring the reader to the books and the
reviews available in the literature for a complete description (see,
for instance, Refs.~\cite{Dreizler} and \cite{booksQP}), before
applying these methods to rutile and anatase TiO$_{\text{2}}$. 

As originally introduced by Hohenberg and Kohn (HK) in
1964~\cite{Hohenberg}, DFT is based on the theorem that the ground
state energy of a system of $N$ interacting electrons in an external
potential $V_{ext}({\bf r})$ is a unique functional of the ground
state electronic density. The Kohn and Sham~\cite{Kohn} formulation
demonstrates how the the HK Theorem may be used in practice, by
self-consistently solving a set of one-electron equations (KS
equations), 
\begin{eqnarray}
\left[-\frac{1}{2}\nabla^2+  V_{eff}^{KS}[\rho({\bf r})]\right]
\phi_i({\bf r})
 = \varepsilon_i^{KS} \phi_i({\bf r}),
\label {eq:KS}
\quad\rho({\bf r}) = \sum_{i}^{N} \vert \phi_i({\bf r}) \vert^2,
\end{eqnarray}
where \(\rho({\bf r})\) is the electronic charge density, \(\phi_i({\bf r})\) are the
non-interacting KS wavefunctions, and $ V_{eff}^{KS}[\rho({\bf r})] 
= V_{ext}({\bf r})  +\ V_H[\rho({\bf r})] + V_{xc}[\rho({\bf r})] $ is the
effective one-electron potential. Here, $V_H$ is the Hartree potential
and $V_{xc}$ is the exchange--correlation potential defined in terms
of the xc-functional $E_{xc}$ as  $V_{xc} = \frac{\delta
  E_{xc}}{\delta \rho(r)}$, which contains all the many-body effects.
$V_{xc}$ is usually calculated in either LDA~\cite{LDA,LDA2} or
GGA~\cite{PBE} approximations.  However, semi-empirical functionals
are also available, called hybrids \cite{PBE0,B3LYP}, which somewhat correct the
deficiencies of LDA and GGA for describing exchange and correlation.
This is accomplished by including an exact exchange contribution. 

Other than the highest occupied molecular orbital (HOMO), KS DFT electronic levels do not correspond to the electronic energies of the many electron system. Indeed, the calculated KS band gaps of semiconductors and insulators severely underestimate the experimental ones. Experimentally, occupied states are accessible by direct photoemission, where an electron is extracted from the system ($N-1$ ground state), while unoccupied states are accessible by inverse photoemission, where one electron is added to the system ($N+1$ ground state).  For isolated systems with a finite number of electrons, the electronic gap may be obtained from the DFT calculated ground-state energies with $N+1$, $N$, and $N-1$ electrons \cite{PerdewLevy}. However, for periodic systems, adding or removing an electron is a non-trivial task.
We therefore summarize in the following a rigorous method for describing
excitations based on the Green's function approach. This method allows
us to properly describe the electronic band structure. Further
information and details about the Green's function approach may be
found elsewhere~\cite{booksQP,booksQP2,HedinLud,strinati}. 

\section{Excited states within Many Body Perturbation Theory}\label{MBPT}

When a bare particle, such as an electron or hole, enters an
interacting system, it perturbs the particles in its vicinity. In
essence, the particle is ``dressed'' by a polarization cloud of the
surrounding particles, becoming a so-called quasi-particle. Using this
concept, it is possible to describe the system through a set of
quasi-particle equations by introducing a non-local, time-dependent,
non-Hermitian operator called the self-energy \(\Sigma\).  This
operator takes into account the interaction of the particle with the
system via 
\begin{equation}
\left[-\frac{1}{2}\nabla^2 + V_{ext} + V_{H}\right]
\Psi_i({\bf r},\omega)
+
\int\Sigma({\bf r},{\bf r}',\omega)\Psi_i({\bf r}',\omega)d{\bf r}'
=E_i(\omega)\Psi_i({\bf r}),
\label{eq:quasi}
\end{equation}
where \(\Psi_i({\bf r})\) is the quasi-particle wavefunction.

Since the operator $\Sigma$ is non-Hermitian, the energies
$E_i(\omega)$ are in general complex, and the imaginary part of
$\Sigma$  is related to the lifetime of the excited
particle~\cite{Rubio}. The most often used approximation to calculate
the self-energy is the so-called $GW$ method. It may be derived as the
first-step iterative solution of the Hedin integral equations (see
Refs.~\cite{GW}, \cite{hyblou}, and \cite{godshsh}), which link the Green's function
$G$, the self-energy  $\Sigma$, the screened Coulomb potential $W$,
the polarization $P$ and the vertex $\Gamma$.  In practice
\ref{eq:quasi} is not usually solved directly, since the KS
wavefunctions are often very similar to the $GW$
ones~\cite{hyblou}. For this reason it is often sufficient to
calculate the quasi-particle (QP) corrections within first-order
perturbation theory ($G_0W_0$) ~\cite{hyblou,godshsh}.  Moreover, the energy
dependence of the self-energy is accounted for by expanding $\Sigma$
in a Taylor series, so that the QP energies 
\(\varepsilon_i^{G_0W_0}\) are then given by 
\begin{eqnarray}
\varepsilon_i^{G_0W_0} &=& \varepsilon_i^{KS} + 
Z_i\langle\phi_i^{KS}|\Sigma(\varepsilon_i^{KS}) - V_{xc}^{KS}|\phi_i^{KS}\rangle,
\end{eqnarray}
where \(Z_i\) are the quasi-particle renomalization factors described
in Refs.~\cite{booksQP} and~\cite{booksQP2}. For a large number
of materials, the  $G_0W_0$ approximation of $\Sigma$ 
works quite well at correcting the KS electronic gap from DFT. 

Concerning optical properties, the physical quantity to be determined
in order to obtain the optical spectra is the macroscopic dielectric
function $\epsilon$$_M$($\omega$). This may be calculated at different
levels of accuracy within a theoretical \emph{ab initio} approach. A
major component in the interpretation of the optical measurements of
reduced dimensional systems are the local-field effects (LFE). These
effects are especially important for inhomogeneous systems. Here, even
long wavelength external perturbations produce microscopic
fluctuations of the electric field, which must be taken into
account. However, LFE are also important for bulk phases such as
anatase and rutile TiO$_{\text{2}}$.  They must be taken into account
in the evaluation of optical absorption, and to calculate the screened
interaction $W$ used in $GW$. The effect becomes increasingly
important when going to lower dimensional systems.   

It is well known~\cite{AdlerWiser} that for inhomogeneous materials
$\epsilon_M(\omega)$ is not simply the average of the corresponding
microscopic quantity, but is related to the inverse of the microscopic
dielectric matrix by 
\begin{eqnarray}
\epsilon_M(\omega) = \lim_{{\bf q} \rightarrow 0}
\frac{1}{\epsilon^{-1}_{{\bf G=0},{\bf G'=0}}({\bf q},\omega) }.
\end{eqnarray}

The microscopic dielectric function may be determined within the
linear response theory~\cite{Lind}, the independent-particle picture
by the random phase approximation (RPA), and using eigenvalues and
eigenvectors of a one-particle scheme  such as DFT or $GW$. There is
also a different formulation which includes LFE in the macroscopic
dielectric function. This becomes useful when the electron--hole
interaction is included in the polarization function. This formulation
allows us to include, via the Bethe-Salpeter equation (BSE), excitonic
and LFE on the same footing.  In so doing, inverting of the
microscopic dielectric matrix is avoided. The complete derivation may
be found in Appendix B of Ref.~\cite{review}. 

So far, in RPA, we have treated the quasi-particles as non-interacting. To take
into account the electron--hole interaction, a higher order vertex
correction needs to be included in the polarization. In other words,
the BSE, which describes the electron--hole pair dynamics, needs to be
solved. As explained in Ref.~\cite{review}, the BSE may be written
as an eigenvalue problem involving the effective two-particle 
Hamiltonian 
\begin{eqnarray}
H_{exc}^{(n_{1},n_{2}),(n_{3},n_{4})}~=~(E_{n_{2}}-E_{n_{1}})\delta
_{n_{1},n_{3}}\delta _{n_{2},n_{4}}-i(f_{n_{2}}-f_{n_{1}})\nonumber\\
\times \int d{\bf r}_{1}\, d{\bf r}_{1}^{\prime }\, d{\bf r}_{2}\,
d{\bf r}
_{2}^{\prime }\ \phi _{n_{1}}({\bf r}_{1})\,\phi _{n_{2}}^{*}({\bf r}
_{1}^{\prime })\,\Xi ({\bf r}_{1},{\bf r}_{1}^{\prime },{\bf r}_{2},{\bf r}
_{2}^{\prime })\,\phi _{n_{3}}^{*}({\bf r}_{2})\,\phi _{n_{4}}({\bf r}
_{2}^{\prime }). \nonumber
\end{eqnarray}
The kernel $\Xi $ contains two contributions: $\bar{v}$, which is the
bare Coulomb interaction with the long range part corresponding to a
vanishing wave vector not included  and $W$,  the attractive screened
Coulomb electron--hole interaction. Using this formalism and
considering only the resonant part of the excitonic
Hamiltonian~\cite{review},  the macroscopic dielectric function may be
expressed as~\cite{review} 
\begin{equation}
\epsilon _{M}(\omega )=1+\lim_{{\bf q}\to 0}v({\bf q})\sum_{\lambda }\frac{
\left| \sum_{v,c;{\bf k}}
\langle v,{\bf k}-{\bf q}|e^{-i{\bf qr}}|c,{\bf k}\rangle
A_{\lambda
}^{(v,c;{\bf k})} \right|^{2}}{(E_{\lambda }-\omega )}.
\label{eps_exc}
\end{equation}
In \ref{eps_exc} the dielectric function, differently from the RPA
approximation, is given by a mixing of single particle transitions
weighted by the excitonic eigenstates $A_{\lambda}$.  These are
obtained by the diagonalization of the excitonic
Hamiltonian. Moreover, the excitation energies in the denominator are
changed from $\epsilon_{c}-\epsilon_{v}$ to $E_{\lambda}$. The
electronic levels are mixed to produce optical transitions, which are
no longer between pairs of independent particles. The excitonic
calculation is in general, from the computational point of view, very
demanding because the matrix to be diagonalized may be very large. The
relevant parameters which determine its size are the number of ${\bf
  k}$-points in the Brillouin zone, and the number of valence and
conduction bands, $N_v$ and $N_c$  respectively, which form the basis
set of pairs of states. 

Calculations performed for insulators and semiconductors show that the
inclusion of the electron--hole Coulomb interaction yields a
near-quantitative agreement with experiment.  This is not only true
below the electronic gaps, where bound excitons are generally formed,
but also above the continuum edge. The same results apply to the
titania-based materials investigated here, as shown in the following
section. 

\section{The bulk phases of TiO$_{\text{2}}$: role of Many Body effects}\label{bulk}

Despite the importance of its surfaces and nanostructures, the most
recent measurements of TiO$_{\text{2}}$'s bulk (see Fig. 1) electronic
and optical properties were performed in the 1960s, with a few
exceptions. Here we aim to review the existing results obtained using
a variety of experimental techniques and \emph{ab initio}
calculations, in order to elucidate the known properties of
TiO$_{\text{2}}$. Previous data will be compared with a complete,
consistent \emph{ab initio} description, which includes many body
effects when describing electronic and optical properties. Most
experimental and theoretical data reported refers to the rutile phase,
while anatase in general has been less studied.  However, anatase has
received more attention recently, due in part to its greater stability
at the nanoscale compared to rutile.   
 
As we will see, while a general agreement seems to exist concerning
the optical absorption edge of these materials, values for basic
electronic properties such as the band gap still have a large degree
of uncertainty. 

The electronic properties of valence states of rutile TiO$_{\text{2}}$
have been investigated experimentally by angle-resolved photoemission
spectroscopy~\cite{Hardman1994}, along the two high symmetry
directions ($\Delta$ and $\Sigma$) in the bulk Brillouin zone. The
valence band of TiO$_{\text{2}}$ consists mainly  of O 2$p$ states
partially hybridized with Ti 3$d$ states. The metal 3$d$ states
constitute the conduction band, with a small amount of mixing with O
2$p$ states. This photoemission data was compared to calculations
performed with both pseudopotentials and linear muffin-tin orbital
(LMTO) methods, which gave a direct gap of 2~eV in both cases. On the
other hand, within the linear combination of atomic orbitals (LCAO)
method, a gap of 3~eV was obtained.  

From the symmetry of the TiO$_6$ octahedrons (see \ref{LFig1}), $d$
states are usually grouped into low energy $t_{2g}$ and high energy
$e_{g}$ sub-bands. It is important to note that, from ultraviolet
photoemission spectroscopy (UPS) data, it has been deduced that  the
electronic gap for rutile is at least 4~eV. This is the observed
binding energy of the first states  below the Fermi energy. This is in
agreement with previous reported data from electron energy loss
spectroscopy~\cite{Eriksen1987}, and from other
UPS~\cite{Heise1992599,Egdell1986835,Kurtz1989178} results.  The
electronic structure of rutile bulk has also been described using
other  experimental techniques, such as electrical
resistivity~\cite{TiO21953},
electroabsorption~\cite{PhysRevLett.17.857,PhysRev.157.700},
photoconductivity and
photoluminescence~\cite{PhysRev.184.979,PhysRevB.51.6842},  X-ray
absorption spectroscopy
(XAS)~\cite{PhysRevB.7.2333,Fischer,PhysRevB.48.2074,Grunes,PhysRevB.41.12366,PhysRevB.40.5715},
resonant Raman spectra~\cite{PhysRevB.51.6842,PhysRevB.46.2024}
photoelectrochemical analysis~\cite{PhysRev.87.876,Kavan} and
UPS~\cite{fleming:033707}.  All of these experiments have provided
many important details of its electronic properties,  in particular
concerning the hybridization between Ti 3$d$ and O 2$p$
states. However,  the electronic band gap,  corresponding to the
difference between the valence band maximum (VBM) and the conduction
band minimum (CBM),  has not been obtained directly from any
experimental data. Although the electronic band gap could be measured
using combined photoemission and inverse photoemission experiments,
such experiments do not appear in the literature.  

The same discussion is valid for the anatase crystalline phase. Even
though there are several XAS measurements concerning its electronic
structure~\cite{PhysRevB.40.5715,sanjines:2945,tang:2042},
photoemission data is completely lacking for anatase. In the absence
of more recent and refined experimental results for rutile, and due to
the lack of results for anatase, we are left with an estimate of 4~eV
for the electronic gap of rutile TiO$_{\text{2}}$.  It is this value
which we shall use as a reference in the following discussion.   

We will now review the experimental results for optical
properties of TiO$_{\text{2}}$.  Such measurements are of great
interest for the photocatalytic and photovoltaic applications of this
material. From the optical absorption spectra of both
phases~\cite{Cardona}, the room temperature optical band gap is found
to be 3.0~eV for rutile, and 3.2~eV for anatase.  

The absorption edge has been investigated in detail for rutile  by
combining absorption, photoluminescence, and Raman scattering
techniques~\cite{Pascual}. These techniques provide a value for the
edge of 3.031~eV associated to a 2$p_{xy}$ exciton, while a lower
energy 1$s$ quadrupolar exciton has been identified below 3~eV. The
first dipole allowed gap is at 4.2~eV~\cite{Pascual} according to the
combined results of these three techniques. Concerning
anatase~\cite{Hosaka199675,HosakaJ},  the optical spectrum have been
recently re-evaluated~\cite{HosakaJ}, confirming the 3.2~eV value for
the edge.  The fine details of anatase's spectrum have also been
recently investigated~\cite{Tang}. Data on the Urbach tail has
revealed that excitons in anatase are self-trapped in the octahedron
of coordination of the titanium atom. This is in contrast to the
rutile phase, where excitons are known to be free due to the different
packing of rutile's octahedra.~\cite{Tang}  

In general, measurements of optical properties can be significantly
affected by the presence of defects, such as oxygen vacancies, and by
phonons. Both defects and phonons will be present in any experimental
sample of the material at finite temperatures. These observations have
to be kept in mind when directly comparing experimental measurements
with the theoretical results presented in the following. Moreover,
there is a general trend in theoretical-computational studies to
compare the theoretical electronic band gaps with the experimental
optical gap
values~\cite{schilfgaarde:226402,Labat1,Muscat,GandugliaPirovano2007219,Diebold200353,schilfgaarde:226402}
derived from the above mentioned experiments.  

It should be remembered that almost by definition, the optical gap is
always smaller than the electronic band gap. This is because the two
types of experiments (photoemission, and optical absorption) provide
information on two different physical quantities. Reverse
photoemission experiments involve a change in the total number of
electrons in the material ($N\rightarrow N+1$), while optical
absorption experiments do not ($N \rightarrow N^*$).  The latter
involves the creation of an electron--hole pair in the material, with
the hole stabilizing the excited electron.  For this reason,
comparison between experimental and theoretical data, and the
resulting discussion, must take into account the proper quantities. 

The theoretical investigations presented in the literature of
TiO$_{\text{2}}$'s structural, electronic, and optical properties are
at varying levels of theory and thus somewhat inconsistent. A
comprehensive description of properties of both phases in the same
theoretical and computational framework is still missing. Here we
present, in a unified description and by treating with the same method
for the two phases, the electronic and optical properties of the two
most stable phases of bulk titania. The \emph{ab intio} calculations
performed yield results~\cite{Chiodo} in quite good agreement with the
few available pieces of experimental data. 

A combination of DFT~\cite{Hohenberg} and many body perturbation
theory (MBPT) methods is a reliable and well established toolkit to
obtain a complete analysis of electronic and optical properties for a
large class of materials and structures. In this DFT + $GW$ + BSE
framework, the properties of  the two bulk phases of TiO$_{\text{2}}$
may be properly analyzed. Their structural and energetic properties
have been calculated~\cite{Chiodo} using  DFT, as a well established
tool for the description of ground state properties.  However, the DFT
gap is, as expected, significantly smaller than the experimental gap,
with the relative positions  of the $s$, $p$, and $d$ levels also
affected by this description. To address this, standard $G_0W_0$
calculations may be applied to obtain  the quasi-particle corrections
to the energy levels, starting from DFT eigenvalues and
eigenfunctions. Finally the electron--hole interaction is included, to
properly describe the optical response of the system. 

The description of ground state properties~\cite{Labat1,muscat2002}
performed in the framework of DFT are generally quite good, with the
structural description of TiO$_{\text{2}}$ systems in reasonable
agreement with experiments~\cite{Chiodo}. The lattice constants are
within 2\% of experiment, while bulk modulii are within  10\% of the
experimental results~\cite{Labat1,muscat2002}, as is often found for
DFT. However, DFT incorrectly predicts the anatase phase to be more
stable than the rutile one, even for a small energy difference,
independently of the xc-functional used~\cite{Labat1}.  

\begin{figure}
\centering
\includegraphics[width=0.8\textwidth]{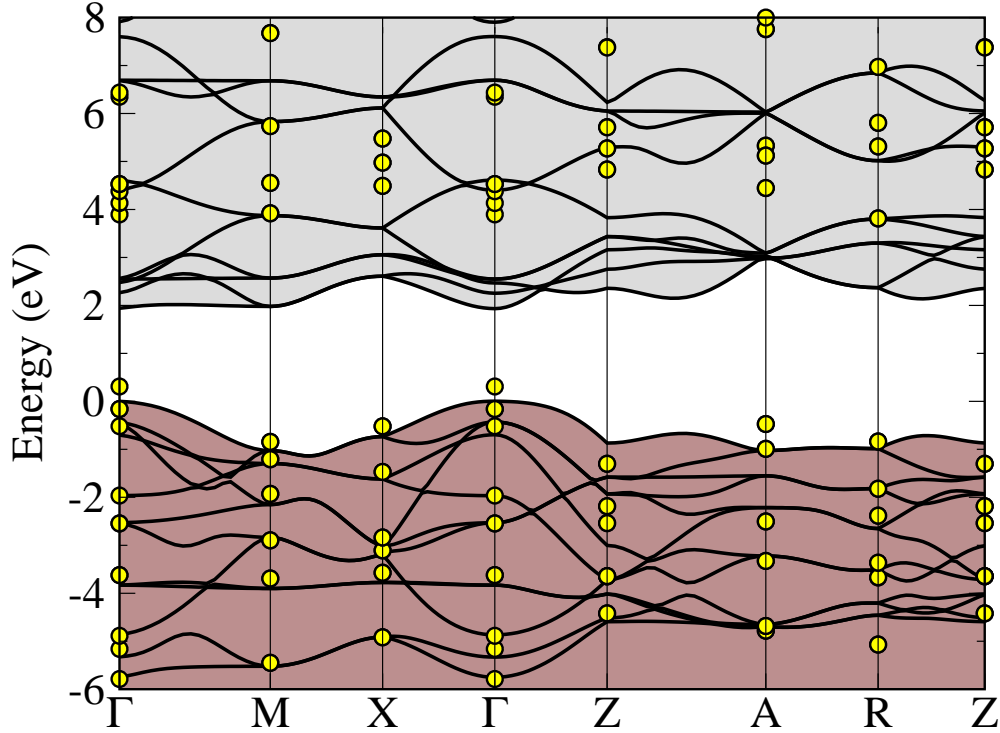}
\caption{Electronic band structure of rutile bulk, along the high symmetry directions of the irreducible Brillouin zone, from GGA calculations ({\textbf{---$\!$---}}). and including the $G_0W_0$ correction ({\color{yellow}{\medbullet}}).}\label{LFig2}
\end{figure}

Even if DFT is not an excited state method, the KS wavefunctions are
often used to evaluate the band structure along the high symmetry
directions (\emph{cf.} \ref{LFig2} for rutile), the density of
states, and the spatial behaviour of wavefunctions involved in the
relevant bonds in the
system~\cite{Chiodo,GW_TiO2_1,PhysRevB.61.7459,Hardman1994}. The KS
electronic gap, corresponding to the difference between the VBM and
the CBM, is 1.93~eV and 2.16~eV for rutile (direct gap) and anatase
(indirect gap), respectively. These are underestimations by almost
2~eV of the available experimental data~\cite{Hardman1994}. However,
the overall behaviour of the band dispersion of KS levels is
reasonable, with valence bands mainly given by O 2$p$ states, and Ti
3$d$ states forming the conduction bands.  

The application of standard $GW$ methods gives gaps of 3.59 and
3.97~eV for rutile and anatase~\cite{Chiodo}, respectively. The value
for rutile is again smaller than the one given by the UPS
estimation~\cite{Hardman1994}, but still close to the experimental
value of 4~eV.

There exist a number of theoretical works, with calculations performed
at different levels of DFT or including MBPT descriptions, for the
electronic gap of rutile and anatase
TiO$_{\text{2}}$~\cite{schilfgaarde:226402,Muscat,schilfgaarde:226402,GW_TiO2_1,PhysRevB.61.7459,PhysRevB.64.184113,GW_TiO2_anatase}. Therefore
in literature it is possible to find for the electronic gap  a quite
large range of possible values, attributed to the gap of titanium
dioxide, which are often erroneously compared with the experimental
optical gap.  

The DFT-GGA values calculated~\cite{Chiodo} are comparable to the ones obtained with a
variety of different DFT approaches, with different functionals, by
using plane waves or localized basis methods, and all-electron or
pseudopotential approaches. Only the hybrid PBE0~\cite{PBE0} and
B3LYP~\cite{B3LYP} xc-functionals give larger values.   

From quasi-particle calculations,  the electronic gap of anatase has
been estimated to be 3.79~eV~\cite{GW_TiO2_anatase} by
$G_0W_0$. However the more refined computational approach, because of
its inclusion of a self-consistent evaluation of $GW$, yields a gap of
3.78~eV for rutile
TiO$_{\text{2}}$.~\cite{schilfgaarde:226402,schilfgaarde:226402} 

\begin{figure}
\centering
\includegraphics[width=\textwidth]{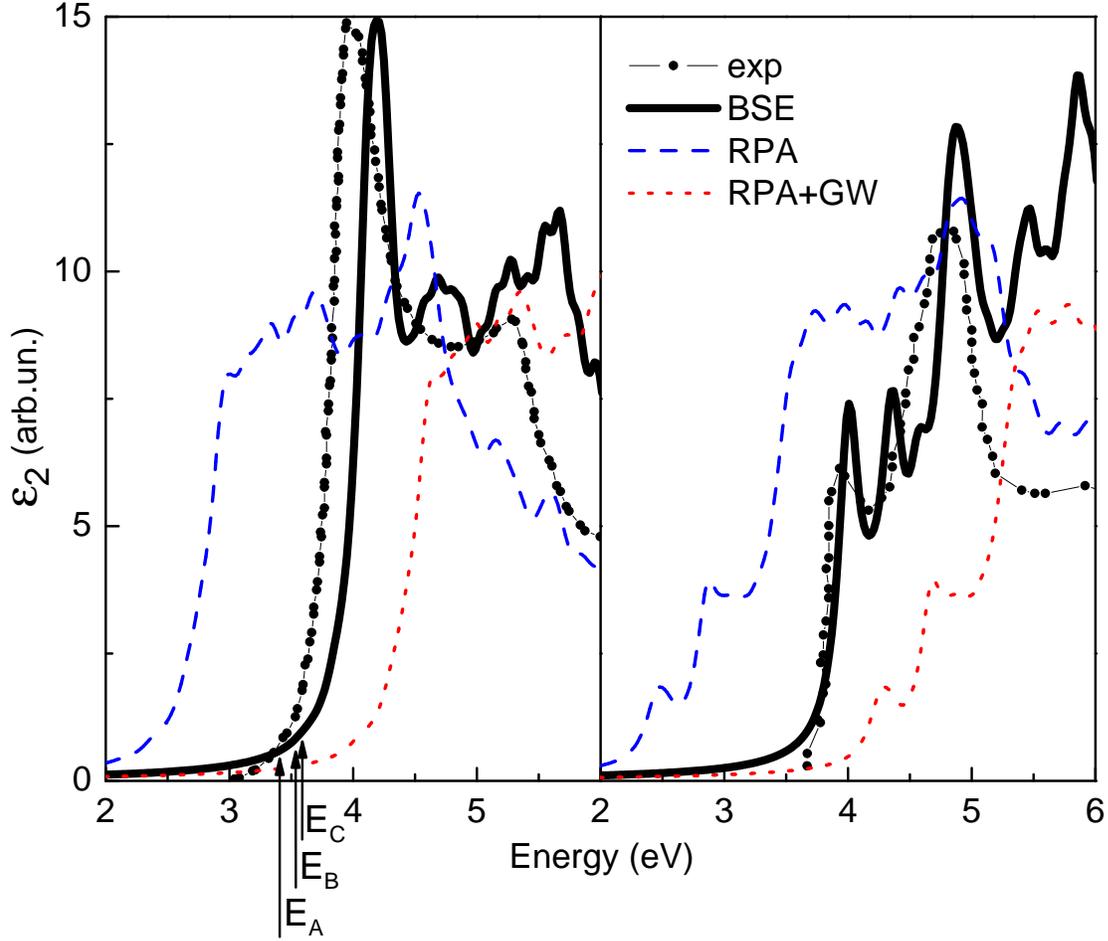}
\caption{Imaginary part of the dielectric constant for rutile (left),
  and anatase (right), in-plane polarization, calculated by  GGA RPA
  ({\color{blue}{\textbf{-- -- --}}}), using $G_0W_0$ on top of GGA
  ({\color{red}{$\cdots\cdots$}}), and via the Bethe-Salpeter equation
  (BSE) ({\color{black}{\textbf{---$\!$---}}}). The experimental
  spectrum ({---~\textbullet~---}) from Refs.~\cite{Cardona} and
  \cite{PhysRevB.61.7459} is also shown for comparison.} 
\label{LFig3}
\end{figure}

Moving to optical properties, and by applying the RPA method to both
KS and QP energies, we obtain spectra (Fig. \ref{LFig3}) that do not
in overall behaviour agree with experiment. Differences are clear both
in absorption edge determination, and in the overall shape of the
spectra. The inclusion of quasi-particle corrections at the $GW$ level
yields a rigid shift of the absorption spectrum, moving the edge at
higher energies, due to the opening of the gap. However, the shape of
the absorption is quite unaffected, because the interaction is still
treated using an independent quasi-particle approximation. A
substantially better agreement may be
obtained~\cite{Chiodo,lawler:205108} by solving the BSE, which takes
into account both many body interactions and excitonic
effects. Indeed, it produces a good description of absorption spectra
and excitons, as shown in Fig. \ref{LFig3} 

The optical absorption spectra calculated for the two phases, with
polarization along the $x$-direction of the unit cells, are provided
in \ref{LFig3}. The spectra given by independent-particle transitions
present two characteristic features.  First, the band edge is
underestimated, due to the electronic gap underestimation in
DFT. Second, the overall shape of the spectrum is, for both phases,
and both orientations, different from the experiment, in the sense
that the oscillator strengths are not correct. The inclusion of the
quasi-particle description, which should improve the electronic gap
description, does not improve the overall shape of the spectrum. The
absorption edge is, however, shifted at higher energies, even higher
than expected from experiments.  The description of optical properties
within the two interacting quasi-particle approximation (by solving
the BSE) definitely improves the results. The absorption edge is now
comparable to the experimental one, with the optical gap estimated
from our calculations in good agreement with the available data.
Further, the shape of the spectrum is now well described, with a
redistribution of transitions at lower energies.  The agreement is
generally good for both phases~\cite{Chiodo}.  

\begin{figure}
\includegraphics[width=\textwidth]{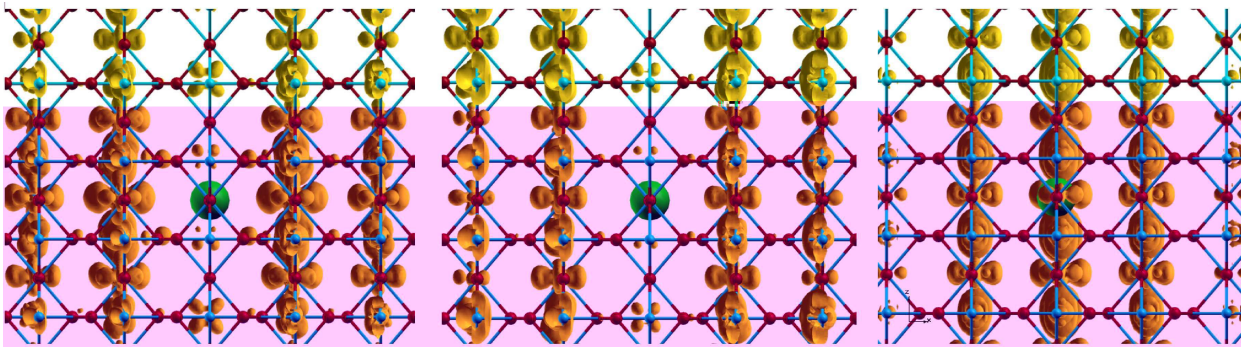}
\caption{Spatial distribution (yellow isosurfaces in arbitrary units) of the partial dark, dark, and optically
  active first three excitons in rutile. The hole position is denoted
  by the light green dot. 
}
\label{LFig4}
\end{figure}

The nature of the exciton is still under debate in TiO$_{\text{2}}$
materials~\cite{Chiodo}. The experimental binding energy is of 4 meV
and some uncertainty exists for the exact determination of optical
edge. Moreover, the exciton is localized in one of the two phases, and
delocalized in the other one, at least based on experimental results.
However, an explanation for this behaviour is so far missing. Refined
measurements \cite{PhysRevB.51.6842} give an exciton of 2$p_{xy}$ character at 3.031~eV, and a
1$s$ quadrupolar exciton at energy lower than 3~eV. From \emph{ab
  initio} calculations with $x$ polarization, two dark (optically inactive) 
or quasi dark excitons are located in rutile at 3.40 and 3.55~eV (denoted by $E_A$,
$E_B$ in \ref{LFig4}).  At the same time, the optically active
exciton, with 4 meV of binding energy, is located at 3.59~eV ($E_C$ in
\ref{LFig4}). The spatial distribution of the first three excitons is
plotted in \ref{LFig4}. The transitions are from O 2$p$ states to Ti 3$d$ states of the triplet $t_{2g}$, 
as expected. While the first two optically forbidden transitions involve Ti atoms farther away from the
excited O atom, the optical active transition involves states of the nearest neighbour Ti atoms. 

In this section we have endeavoured to clarify though the application
of a consistent description, the properties of the two main
crystalline phases of titania. Particular attention has been taken to
how the inclusion of a proper description of exchange and correlation
effects can improve the description of both electronic and optical
properties of TiO$_{\text{2}}$.  

Now that bulk properties are known, from a theoretical point of view,
at the level that the state-of-the-art \emph{ab initio} techniques
allow us to reach, we can attempt to describe how quantum confinement
induced by reduced dimensionality, and doping both effect the
properties of TiO$_{\text{2}}$. Our final aim is to demonstrate how we
may tune the electronic and optical gap of nanostructures for
photocatalytic and photovoltaic applications. 

\section{Using nanostructure to tune the energy gap}\label{nanostructure}

Having described in the previous section the electronic structure of
the two bulk TiO$_{\text{2}}$ phases, we will now turn our attention
to the influence of nanostructure on the energy gap.  This is an area
which has received considerable attention in recent
years~\cite{Structures,ExpTiO2clusters,TiO2NTs1,TiO2NTs2,Delaminated,TiO2},
in part due to the inherently high surface to volume ratio of
nanostructures.  It is hoped that this will allow materials with
shorter quasi-particle lifetimes to function effectively for
photocatalytic activities, since excitons are essentially formed at
the material's surface.  However, this advantage is partly countered
by quantum confinement effects, which tend to increase the energy gap
in nanostructures.  These competing factors make the accurate
theoretical determination of energy gaps in nanostructured materials a
thing of great interest. 

\begin{figure}
\centering
\includegraphics[width=\textwidth]{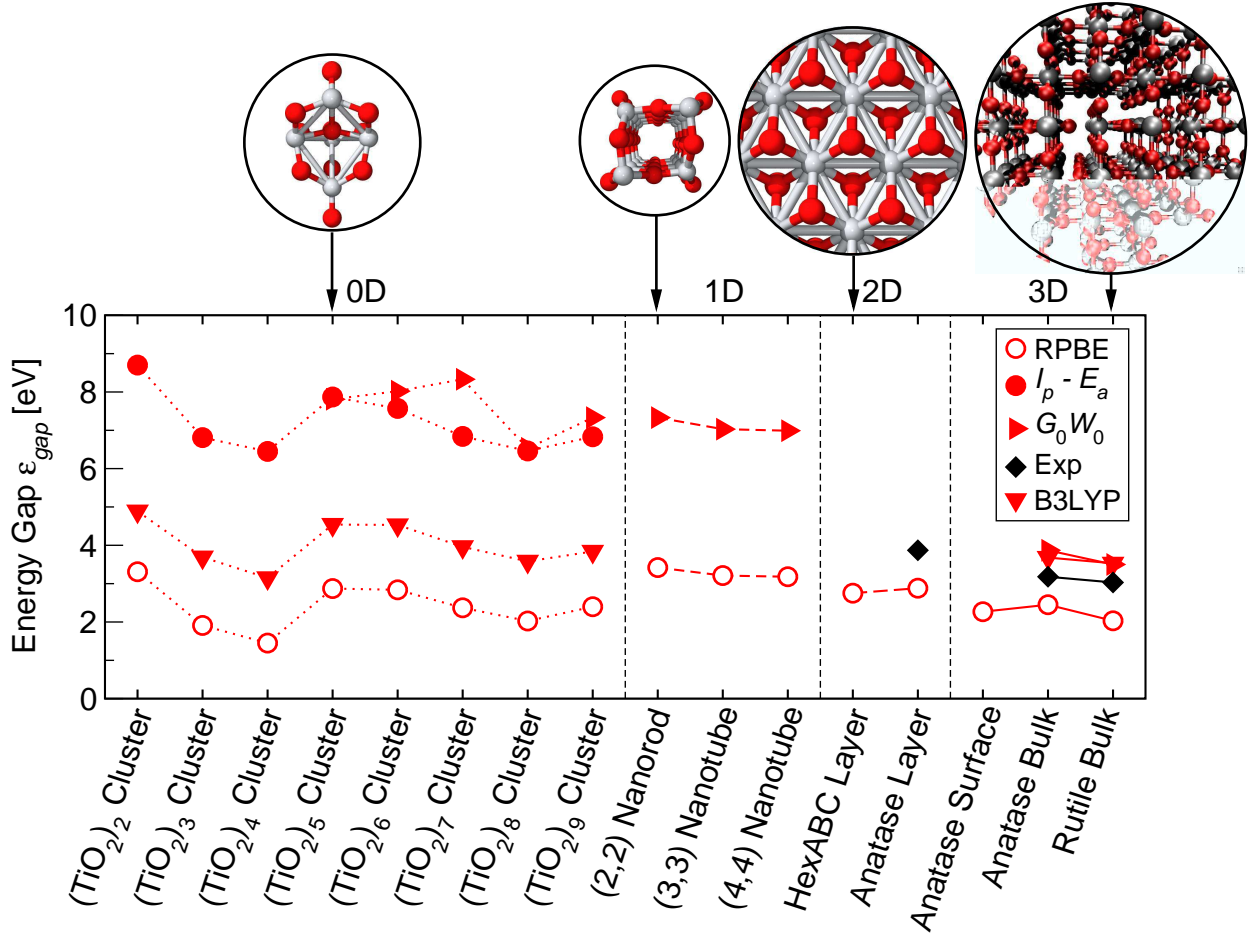}
\caption{Energy gap $\varepsilon_{gap}$  in~eV versus TiO${}_{\text{2}}$
  structure for 0D (TiO${}_{\text{2}}$)${}_n$ clusters ($n\leq {9}$), 1D
  TiO${}_{\text{2}}$ (2,2) nanorods, (3,3) nanotubes, (4,4) nanotubes, 2D
  HexABC and anatase layers, and 3D anatase surface, anatase bulk, and
  rutile bulk phases. DFT calculations using the highest occupied and
  lowest unoccupied state gaps with the standard GGA RPBE
  xc-functional ({\color{red}{$\medcirc$}}) and the hybrid B3LYP
  xc-functional ({\color{red}{$\blacktriangledown$}}), are compared
  with DFT $I_p - E_a$ calculations ({\color{red}{\medbullet}}),
  $G_0W_0$ quasi-particle calculations
  ({\color{red}{$\blacktriangleright$}}), and experimental results
  (\Diamondblack)~\cite{TiO2,Structures,GW_TiO2_anatase,N-TiO2NTarraydoping,TiO2Rev1}. Schematics
  of representative structures for each dimensionality are shown above
  and taken from Ref.~\cite{TiO2}. 
}\label{Undoped} 
\end{figure}

However, as shown in the previous section, standard DFT calculations
tend to underestimate electronic band gaps for bulk TiO$_{\text{2}}$
by approximately 2~eV, due in part to self-interaction
errors~\cite{BandGapErrors,NewXCFuncs}.  These errors arise from an
incomplete cancellation of the electron's Coulomb potential in the
exchange-correlation (xc)-functional. 

This may be partially addressed by the use of hybrid functionals such
as B3LYP~\cite{B3LYP}, which generally seem to improve band gaps for
bulk systems~\cite{Muscat, DeAngelis,DiValentin}. However, such
calculations are computationally more expensive, due to the added
dependence of the xc-functional on the electron's wavefunction.
Moreover, B3LYP calculations for TiO$_{\text{2}}$ clusters largely
underestimate the gap relative to the more reliable difference between
standard DFT calculated ionization potential $I_p$ and electron
affinity $E_a$.  For isolated systems such as clusters, the needed
energetics of charged species are quantitatively described by standard
DFT.  Also, B3LYP and
RPBE~\cite{RPBE} calculations provide the same qualitative description
of the \emph{trends} in the energy gaps for TiO$_{\text{2}}$, as seen
in \ref{Undoped}.  Another methodology is thus needed to describe the
band gaps of periodic systems.    

$G_0W_0$ is probably the most successful and generally applicable method for
calculating quasi-particle gaps.  For clusters it agrees well with
$I_p - E_a$, and it has been shows to produce reliable results for
bulk phases. On the other hand, $G_0W_0$ calculations describe an
$N\rightarrow N+1$ transition where the number of charges is not
conserved, rather than the electron--hole pair induced by
photoabsorption, which is a neutral process.  Indeed, a description in
terms of electronic gap cannot be compared with, or provide direct
information on the optical gap, which is the most investigated
quantity, due to its critical importance for photocatalytic
processes. 

\ref{Undoped} shows that for both 3D and 2D systems, RPBE gaps
underestimate the experimental optical gaps by approximately 1~eV.  For 1D
and 0D systems, there is a much larger difference of about 4~eV and
5~eV respectively, between the RPBE gaps and the $I_p-E_a$ and
$G_0W_0$ results.  This increasing disparity may be attributed to the
greater quantum confinement and charge localization in the 1D and 0D
systems, which yield greater self-interaction effects.  The B3LYP gaps
also tend to underestimate this effect, simply increasing the RPBE
energy gaps for both 0D and 3D systems by about 1.4~eV.     

On the other hand, the RPBE gaps reproduce qualitatively the
structural dependence of the $I_p - E_a$, $G_0W_0$, and experimental
results for a given dimensionality, up to a constant shift.  This is
true even for 3D bulk systems, where standard DFT does not correctly
predict rutile to be the most stable structure~\cite{Labat1}. 

To summarize, quantum confinement effects seem to increase the energy
gap significantly for both 0D and 1D systems, while 2D and 3D systems
may be much less affected.  This suggests that 2D laminar structures
are viable candidates for reducing the minimal quasi-particle, while
leaving the band gap nearly unchanged.  However, a more accurate
description of the photoabsorption properties of these novel
nanostructures, perhaps using the methodologies recently applied to
bulk TiO$_{\text{2}}$~\cite{Chiodo}, still remains to be found. 

\section{Influence of boron and nitrogen doping on TiO$_{\text{2}}$'s energy gap}\label{doping}

The doping of TiO$_{\text{2}}$ nanostructures has received much recent
attention as a possible means for effectively tuning
TiO$_{\text{2}}$'s band gap into the visible
range~\cite{TiO2,JACS-suil,N-TiO2NTarraydoping,N-TiO2doping,N-TiO2NTdoping,N-TiO2NTdoping2,Nambu,Graciani,TiO2PRL,zhu:226401}.
Recent experiments suggest substituting oxygen by boron or nitrogen in
the bulk introduces mid-gap states, allowing lower energy excitations.
However, to model such systems effectively requires large supercells,
both to properly describe the experimental doping fractions of
\(\lesssim\) 10\%, and to ensure dopant--dopant interactions are
minimized.   

\begin{figure}
\centering
\includegraphics[width=\textwidth]{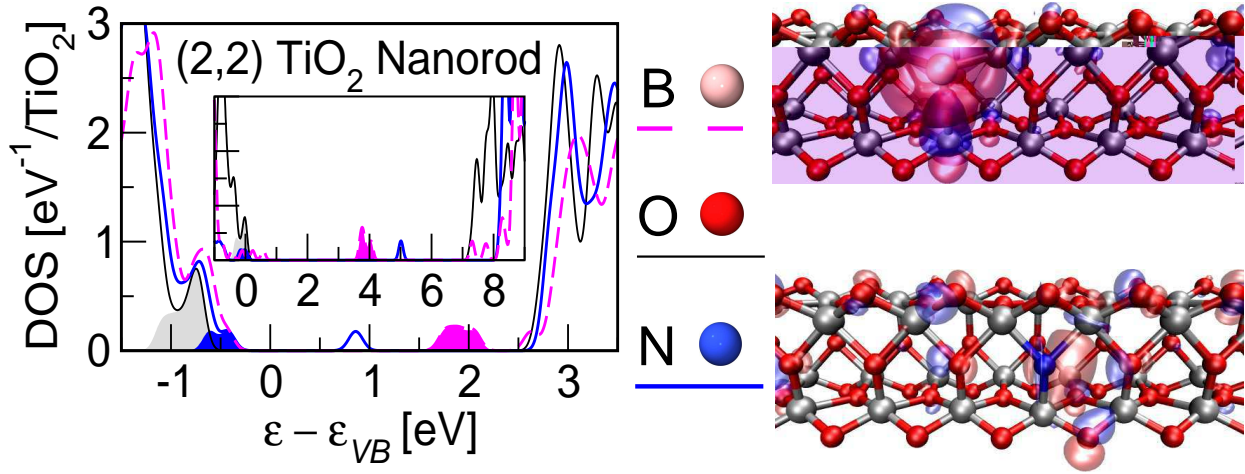}
\caption{Total density of states (DOS) in~eV$^{-1}$ per
  TiO$_{\text{2}}$ functional unit vs.~energy \(\varepsilon\) in~eV
  for undoped  ({\color{black}{---$\!$---}}), boron doped
  ({\color{magenta}{\textbf{-- -- --}}}) and nitrogen doped
  ({\color{blue}{\textbf{---$\!$---}}}) (2,2) TiO$_{\text{2}}$
  nanorods from standard DFT GGA RPBE xc-functional and (inset)
  $G_0W_0$ quasi-particle calculations.  The highest occupied states
  for boron and nitrogen doped (2,2) TiO$_{{2}}$ nanorods are depicted
  by isosurfaces of $\pm$0.05$e/$\AA$^{{3}}$ in the structure diagrams
  to the left.}\label{DOS} 
\end{figure}

\ref{DOS} shows the DFT calculated DOS and structures for the most
stable boron doped and nitrogen doped TiO${}_{\text{2}}$ (2,2)
nanorods.  The highest occupied state is also shown as isosurfaces of
$\pm$0.05$e/$\AA${}^{{3}}$ in the side views of the doped structures.    

As with TiO${}_{\text{2}}$ clusters, the influence of boron dopants on
TiO${}_{\text{2}}$ nanorods may be understood in terms of boron's weak
electronegativity, especially when compared with the strongly
electronegative oxygen.  Boron prefers to occupy oxygen sites which
are 2-fold coordinated to neighbouring titanium atoms. However, as
with the 0D clusters, boron's relatively electropositive  character
induces significant structural changes in the 1D structures, creating
a stronger third bond to a neighbouring three-fold coordinated oxygen
via an oxygen dislocation, as shown in \ref{DOS}. This yields three
occupied mid-gap states localized on the boron dopant, which overlap
both the valence band O 2$p_\pi$ and conduction band  Ti 3$d_{x y}$
states, as shown in \ref{DOS}. Boron dopants thus yield donor states
near the conduction band, which may be photocatalytically active in
the visible region.  However, the quantum confinement inherent in
these 1D structures may stretch these gaps, as found for the $G_0W_0$
calculated DOS shown in the inset of \ref{DOS}.  

On the other hand, nitrogen dopants prefer to occupy oxygen sites
which are 3-fold coordinated to Ti, as was previously found for the
rutile TiO${}_{\text{2}}$ surface~\cite{Nambu, Graciani}.  This yields one
occupied state at the top of the valence band and one unoccupied
mid-gap state in the same spin channel. Both states are localized on
the nitrogen dopant but overlap the valence band O 2$p_\pi$ states, as
shown in \ref{DOS}.  Nitrogen dopants thus act as acceptors,
providing localized states well above the valence band, as is also
found for the $G_0W_0$ calculated DOS shown in the inset of \ref{DOS}. 

Although nitrogen dopants act as acceptors in TiO${}_{\text{2}}$ 1D
structures, such large gaps between the valence band and the
unoccupied mid-gap states would not yield $p$-type semiconductors.
This may be attributed to the substantial quantum confinement in these
1D structures.  However, for 2D and 3D systems, it is possible
to produce both $p$-type and $n$-type classical semiconductors, as
discussed in Ref.~\cite{TiO2}.  This has recently been shown
experimentally in Ref.~\cite{zhu:226401}, where co-doping of
anatase TiO$_{\text{2}}$ with nitrogen and chromium was found to
improve the localization of the acceptor states, and reduce the
effective optical gap. By replacing both Ti$^{\text{4+}}$ and
O$^{\text{2-}}$ atoms with dopants in the same TiO$_6$ octahedral, it
should be possible to ``tune'' the optical band gap to a much finer
degree.   

\begin{figure}
\centering
\includegraphics[width=\textwidth]{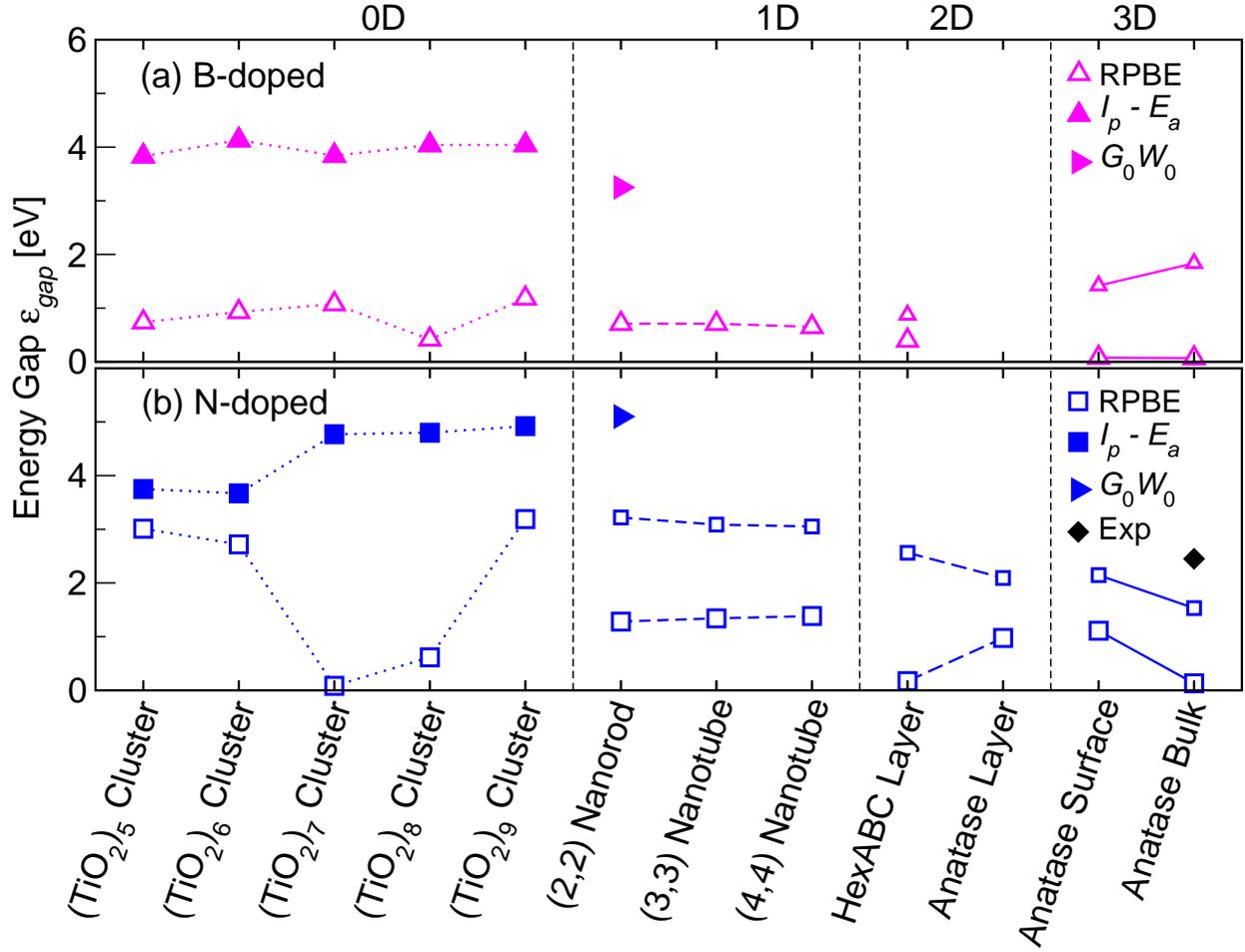}
\caption{Influence of doping on the energy gap $\varepsilon_{gap}$
  in~eV versus TiO${}_{\text{2}}$ structure for 0D
  (TiO${}_{\text{2}}$)${}_n$ clusters ($n\leq {9}$), 1D
  TiO${}_{\text{2}}$ (2,2) nanorods, (3,3) nanotubes, (4,4) nanotubes,
  2D HexABC and anatase layers, and 3D anatase surface and bulk. The
  energy gaps for (a) boron doped systems from DFT calculations using
  the highest occupied and lowest unoccupied state gaps with the
  standard GGA RPBE xc-functional ({\color{magenta}{$\vartriangle$}})
  are compared with DFT $I_p - E_a$ calculations
  ({\color{magenta}{$\blacktriangle$}}), and $G_0W_0$ quasi-particle
  calculations ({\color{magenta}{$\blacktriangleright$}}), and (b)
  nitrogen doped systems from DFT calculations using the highest
  occupied and lowest unoccupied state gaps with the standard GGA RPBE
  xc-functional ({\color{blue}{$\boxempty$}}) are compared with DFT
  $I_p - E_a$ calculations ({\color{blue}{$\blacksquare$}}), and
  $G_0W_0$ quasi-particle calculations
  ({\color{blue}{$\blacktriangleright$}}), and experimental
  (\Diamondblack)
  results~\cite{TiO2,Structures,GW_TiO2_anatase,N-TiO2NTarraydoping,TiO2Rev1}. Small
  open symbols denote transitions between highest fully occupied
  states and the conduction band.}\label{Doped}  
\end{figure}

Whether calculated using RPBE, $I_p-E_a$ or $G_0W_0$, the energy gaps for
both boron and nitrogen doped TiO$_{\text{2}}$ nanostructures are generally
narrowed, as shown in \ref{Doped}(a) and (b).  However, for nitrogen
doped (TiO${}_{\text{2}}$)$_n$ clusters where nitrogen acts as an acceptor ($n
= 5, 6, 9$), the energy gap is actually \emph{increased} when spin is
conserved, compared to the undoped clusters in RPBE.  This effect is
not properly described by the $N\rightarrow N+1$ transitions of
$I_p-E_a$, for which spin is not conserved for these nitrogen doped
clusters.  On the other hand, when nitrogen acts as a donor ($n=7,8$)
the smallest gap between energy levels does conserve spin.    

The boron doped TiO$_{2}$ nanorods and nanotubes have perhaps the most
promising energy gap results of the TiO$_{\text{2}}$ structures, as
seen in \ref{Doped}(a).  Boron dopants introduce in the nanorods
localized occupied states near the conduction band edge in both RPBE
(\emph{cf.} \ref{DOS}) and $G_0W_0$ (\emph{cf.} inset of \ref{DOS})
calculations.  On the other hand, nitrogen doping of nanorods
introduces well defined mid-gap states, as shown in \ref{DOS}.
However, to perform water dissociation, the energy of the excited
electron must be above that for hydrogen evolution, with respect to
the vacuum level.  This is not the case for such a mid-gap state.
This opens the possibility of a second excitation from the mid-gap
state to the conduction band. However, the cross section for such an
excitation may be rather low.  

For boron doping of 2D and 3D structures, the highest occupied state
donates its electron almost entirely to the conduction band, yielding
an $n$-type semiconductor.  Thus at very low temperatures, the RPBE
band gap is very small.  The same is true for $n$-type nitrogen doped
bulk anatase.  For these reasons the energy gap between the highest
fully occupied state and the conduction band, which may be more
relevant for photoabsorption, is also shown.  These RPBE gaps are
still generally smaller than those for their undoped TiO$_{\text{2}}$
counterparts, shown in \ref{Undoped}.  

In summary, for both boron and nitrogen doped clusters we find RPBE
gaps differ from $I_p - E_a$  by about 3~eV, while for nitrogen doped
anatase the RPBE gap differs from experiment by about 0.6~eV.  Given
the common shift of 1~eV for undoped 2D and 3D structures, this
suggests that both boron and nitrogen doped 2D TiO${}_{\text{2}}$
structures are promising candidates for photocatalysis.  Further, the
boron and nitrogen doped 1D nanotube results also  warrant further
experimental investigation.  

\section{Solar cells from TiO$_{\text{2}}$ nanostructures: dye-sensitized solar cells }\label{cells}

TiO$_{\text{2}}$ is so far the most widely used solid material in the
development of solar cell devices based on hybrid
architectures\cite{Gratzel2001,Gratzel2003}. In these devices, the
dye, synthetic or organic, absorbs light, and electrons excited by the
phonons are injected into the underlying oxide nanostructure. The
hybrid system must therefore satisfy several requirements: (1) a
proper absorption range for the dye, (2) a fast charge transfer in the
oxide, (3) a slow back-transfer process, and (4) an easy collection
and conduction of electrons in the oxide.  

While absorption properties may be easily tuned at the chemical level
by changing or adding functional groups, a critical point is to
understand, and therefore control, the process at the
interface. Indeed, simply having a good energy level alignment is not
sufficient because the fast electron injection process is
dynamical. Since the experimental characterization of complex systems
(networks of nanostructures, with adsorbed dyes, and in solution) is
quite complicated, the theoretical description of such systems can be
of fundamental importance in unravelling the processes governing the
behaviour of dye-sensitized solar cells (DSSC).  

\begin{figure}
\centering
\includegraphics[width=0.8\textwidth]{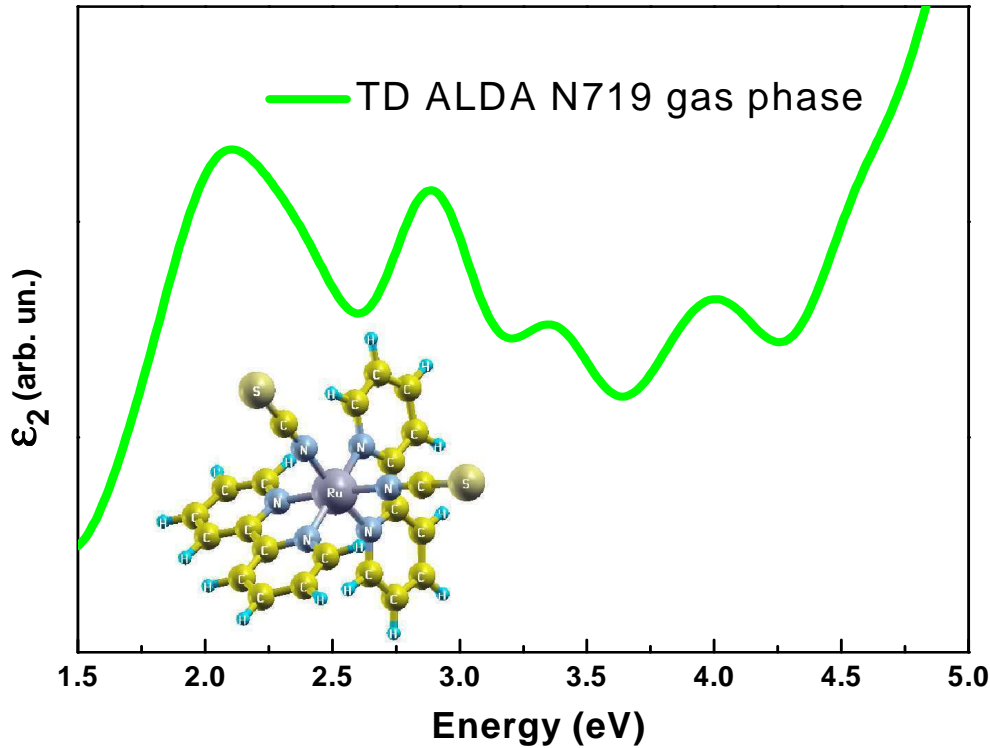}
\caption{Absorption spectrum calculated by TD Adiabatic LDA of the Ru-dye N719, 
whose structure is shown in the inset.
}\label{N719}
\end{figure}

The most popular technique for studying these dynamical processes is
time dependent DFT (TDDFT). For a detailed discussion of the success
and possible limitations of this method when applied to hybrid
systems, we refer the reader to Ref.~\cite{tddftbook}. TDDFT is a
generalization of DFT which allows us to directly describe excited
states.  For example, it has previously been used successfully to
calculate the optical absorption of large organic molecules
(\ref{N719}), such as Ru-dyes, or indolines. This same technique has also been applied recently to
hybrid systems used for photovoltaic applications. We want here to
highlight that the charge injection transfer has also been modelled
for molecules adsorbed on TiO$_{\text{2}}$
clusters\cite{deangelis2004,deangelis2005,deangelis2005-2,deangelis2007,deangelis2008,prezhdo2005,prezhdo2007,prezhdo2007-2,prezhdo2008,prezhdo2008-2},
giving estimations as fast as 8 fs for the injection
time~\cite{prezhdo2005}. TDDFT is therefore a powerful tool to
understand dynamical electronic effects in systems as large and
complex as hybrid organic-oxide solar cells.

\section{Conclusion}

In this chapter we showed how an oxide of predominant importance in
nanotechnological and environmental fields, such as photocatalysis and
photovoltaics, can be successfully investigates by state-of-the-art
\emph{ab initio} techniques. The optical and excitonic properties of
the bulk phases can be properly described by MBPT techniques. We also
showed how the electronic properties of TiO$_{\text{2}}$ may be
``tailored'' using nanostructural changes in combination with boron
and nitrogen doping.  While boron doping tends to produce smaller band
gap $n$-type semiconductors, nitrogen doping produces $p$-type or
$n$-type semiconductors depending on whether or not nearby oxygen
atoms occupy surface sites.  This suggests that a $p$-type
TiO${}_{\text{2}}$ semiconductor may be produced using nitrogen doping
in conjunction with surface confinement at the nanoscale.  We also
showed that it has been proved how, in the field of photovoltaics,
TDDFT is a powerful tool to understand the mechanism of charge
injection at organic-inorganic interfaces. 

\acknowledgments
We acknowledge funding by ``Grupos Consolidados UPV/EHU del Gobierno
Vasco'' (IT-319-07), the European Community through  e-I3 ETSF project
(Contract Number 211956). We acknowledge support by the Barcelona
Supercomputing Center, ``Red Espanola de Supercomputacion'', SGIker
ARINA (UPV/EHU) and Transnational Access Programme
HPC-Europe++. J.M. G-L. acknowledges funding from the Spanish MEC
through the `Juan de la Cierva' program. L.C.  acknowledges funding
from UPV/EHU through the `Ayudas de Especializaci{\'{o}}n para
Investigadores Doctores' program. We also thank H. Petek, S. Ossicini,
M. Wanko, M. Piacenza, and M. Palummo for helpful discussions. 


\begin{thebibliography}{111}
\providecommand{\natexlab}[1]{#1}
\providecommand{\url}[1]{\texttt{#1}}
\expandafter\ifx\csname urlstyle\endcsname\relax
  \providecommand{\doi}[1]{doi: #1}\else
  \providecommand{\doi}{doi: \begingroup \urlstyle{rm}\Url}\fi

\bibitem{Gratzel2001}
M.~Gr{\"{a}}tzel, Photoelectrochemical cells, \emph{Nature} {\bf 414},
  \penalty0 338--344  (2001).

\bibitem{FirstPhotocatalysis}
A.~Heller, Hydrogen-evolving solar cells, \emph{Science} {\bf 223}, \penalty0
  1141--1148  (1984).

\bibitem{Hoffmann1995}
M.~R. Hoffmann, S.~T. Martin, W.~Y. Choi, and D.~W. Bahnmann, Environmental
  applications of semiconductor photocatalysis, \emph{Chem. Rev.} {\bf 95},
  \penalty0 69--96  (1995).

\bibitem{C-TiO2doping}
S.~U.~M. Khan, M.~Al-Shahry, and W.~B.~I. Jr., Efficient photochemical water
  splitting by a chemically modified n-{TiO$_{{2}}$}, \emph{Science} {\bf
  297}, \penalty0 2243--2245  (2002).

\bibitem{TiO2PRL}
Y.~Gai, J.~Li, S.-S. Li, J.-B. Xia, and S.-H. Wei, Design of narrow-gap
  {TiO$_2$}: A passivated codoping approach for enhanced photoelectrochemical
  activity, \emph{Phys. Rev. Lett.} {\bf 102}, \penalty0 036402--1--4  (2009).

\bibitem{Structures}
Z.-W. Qu and G.-J. Kroes, Theoretical study of the electronic structure and
  stability of titanium dioxide clusters ({TiO$_{{2}}$})$_n$ with $n$ = 1--9,
  \emph{J. Phys. Chem. B} {\bf 110}, \penalty0 8998--9007  (2006).

\bibitem{ExpTiO2clusters}
H.-J. Zhai and L.-S. Wang, Probing the electronic structure and band gap
  eveolition of titanium oxide clusters ({TiO$_2$})$_n^-$ ($n$ = 1--10) using
  photoelectron spectroscopy, \emph{J. Am. Chem. Soc.} {\bf 129}, \penalty0
  3022--3026  (2007).

\bibitem{TiO2NTs1}
H.~Imai, Y.~Takei, K.~Shimizu, M.~Matsuda, and H.~Hirashima, Direct preparation
  of anatase {TiO}$_{{2}}$ nanotubes in porous alumina membranes, \emph{J.
  Mater. Chem.} {\bf 9}, \penalty0 2971--2972  (1999).

\bibitem{TiO2NTs2}
T.~Kasuga, M.~Hiramatsu, A.~Hoson, T.~Sekino, and K.~Niihara, Titania nanotubes
  prepared by chemical processing, \emph{Adv. Mater.} {\bf 11}, \penalty0
  1307--1311  (1999).

\bibitem{Delaminated}
G.~Mogilevsky, Q.~Chen, H.~Kulkarni, A.~Kleinhammes, W.~M. Mullins, and Y.~Wu,
  Layered nanostructures of delaminated anatase: Nanosheets and nanotubes,
  \emph{J. Phys. Chem. C} {\bf 112}, \penalty0 3239--3246  (2008).

\bibitem{TiO2}
D.~J. Mowbray, J.~I. Mart{\'{\i}}nez, J.~M. {Garc{\'{\i}}a Lastra}, K.~S.
  Thygesen, and K.~W. Jacobsen, Stability and electronic properties of
  {TiO}$_{{2}}$ nanostructures with and without {B} and {N} doping, \emph{J.
  Phys. Chem. C} {\bf 113}, \penalty0 12301--12308  (2009).

\bibitem{JACS-suil}
S.~In, A.~Orlov, R.~Berg, F.~Carc{\'{\i}}a, S.~Pedrosa-Jimenez, M.~S. Tikhov,
  D.~S. Wright, and R.~M. Lambert, Effective visible light-activated {B}-doped
  and {B,N}-codoped {TiO$_{{2}}$} photocatalysts, \emph{J. Am. Chem. Soc.} {\bf
  129}, \penalty0 13790--13791  (2007).

\bibitem{N-TiO2NTarraydoping}
G.~Liu, F.~Li, D.-W. Wang, D.-M. Tang, C.~Liu, X.~Ma, G.~Q. Lu, and H.-M.
  Cheng, Electron field emission of a nitrogen-doped {TiO$_{{2}}$} nanotube
  array, \emph{Nanotechnology} {\bf 19}, \penalty0 025606--025611  (2008).

\bibitem{N-TiO2doping}
R.~Asahi, T.~Morikawa, T.~Ohwaki, K.~Akoki, and Y.~Taga, Visible-light
  photocatalysis in nitrogen-doped titanium oxides, \emph{Science} {\bf 298},
  \penalty0 269--271  (2001).

\bibitem{N-TiO2NTdoping}
A.~Ghicov, J.~M. Macak, H.~Tsuchiya, J.~Kunze, V.~H{\ae}ublein, S.~Kleber, and
  P.~Schmuki, {TiO}$_{{2}}$ nanotube layers: Dose effects during nitrogen
  doping by ion implantation, \emph{Chem. Phys. Lett.} {\bf 419}, \penalty0
  426--429  (2006).

\bibitem{N-TiO2NTdoping2}
Y.~Chen, S.~Zhang, Y.~Yu, H.~Wu, S.~Wang, B.~Zhu, W.~Huang, and S.~Wu,
  Synthesis, characterization, and photocatalytic activity of {N}-doped
  {TiO$_{{2}}$} nanotubes, \emph{J. Disp. Sci. Tech.} {\bf 29}, \penalty0
  245--249  (2008).

\bibitem{Nambu}
A.~Nambu, J.~Graciani, J.~A. Rodriguez, Q.~Wu, E.~Fujita, and J.~F. Sanz, N
  doping of {TiO$_2$}(110): Photoemission and density-functional studies,
  \emph{J. Chem. Phys.} {\bf 125}, \penalty0 094706--1--8  (2006).

\bibitem{Graciani}
J.~Graciani, L.~J. {\'{A}}lvarez, J.~A. Rodriguez, and J.~F. Sanz, N doping of
  rutile {TiO$_2$}(110) surface. a theoretical {DFT} study, \emph{J. Phys.
  Chem. C} {\bf 112}, \penalty0 2624--2631  (2008).

\bibitem{zhu:226401}
W.~Zhu, X.~Qiu, V.~Iancu, X.-Q. Chen, H.~Pan, W.~Wang, N.~M. Dimitrijevic,
  T.~Rajh, H.~M.~M. III, M.~P. Paranthaman, G.~M. Stocks, H.~H. Weitering,
  B.~Gu, G.~Eres, and Z.~Zhang, Band gap narrowing of titanium oxide
  semiconductors by noncompensated anion-cation codoping for enhanced
  visible-light photoactivity, \emph{Phys. Rev. Lett.} {\bf 103}, \penalty0
  226401--1--4  (2009).

\bibitem{TiO2Rev1}
D.~V. Bavykin, J.~M. Friedrich, and F.~C. Walsh, Protonates titanates and
  {TiO$_{2}$} nanostructured materials: Synthesis, properties, and
  applications, \emph{Adv. Mater.} {\bf 18}, \penalty0 2807--2824  (2006).

\bibitem{Turner}
O.~Khaselev and J.~A. Turner, A monolithic photovoltaic-photoelectrochemical
  device for hydrogen production via water splitting, \emph{Science} {\bf
  280}, \penalty0 425--427  (1998).

\bibitem{Nazeeruddin1}
M.~K. Nazeeruddin, A.~Kay, I.~Rodicio, R.~Humphry-Baker, E.~M{\"{u}}ller,
  P.~Liska, N.~Vlachopoulos, and M.~Gr{\"{a}}tzel, Conversion of light to
  electricity by
  {\emph{cis}}-{X$_2$Bis}(2,2'-bipyridyl-4,4'-dicarboxylate)ruthenium(ii)
  charge-transfer sensitizers ({X = Cl$^-$}, {Br$^-$}, {I$^-$}, {CN$^-$}, and
  {SCN$^-$}) on nanocrystalline {TiO$_2$} electrodes, \emph{J. Am. Chem. Soc.}
  {\bf 115}, \penalty0 6382--–6390  (1993).

\bibitem{Nazeeruddin2}
P.~Wang, S.~M. Zakeeruddin, J.~E. Moser, R.~Humphry-Baker, P.~Comte,
  V.~Aranyos, A.~Hagfeldt, M.~K. Nazeeruddin, and M.~Gr{\"{a}}tzel, Stable new
  sensitizer with improved light harvesting for nanocrystalline dye-sensitized
  solar cells, \emph{Adv. Mater.} {\bf 16}, \penalty0 1806--1811  (2004).

\bibitem{Mao}
X.~Chen and S.~S. Mao, Titanium dioxide nanomaterials: Synthesis, properties,
  modifications, and applications, \emph{Chem. Rev.} {\bf 107}, \penalty0
  2891–--2959  (2007).

\bibitem{Meyer}
D.~F. Watson and G.~J. Meyer, Electron injection at dye-sensitized
  semiconductor electrodes, \emph{Ann. Rev. Phys. Chem.} {\bf 56}, \penalty0
  119--156  (2005).

\bibitem{Jih-Jen}
J.-J. Wu, G.-R. Chen, C.-C. Lu, W.-T. Wu, , and J.-S. Chen, Performance and
  electron transport properties of {TiO$_2$} nanocomposite dye-sensitized solar
  cells, \emph{Nanotechnology} {\bf 19}, \penalty0 105702--105708  (2008).

\bibitem{Kang}
S.~H. Kang, S.-H. Choi, M.-S. Kang, J.-Y. Kim, H.-S. Kim, T.~Hyeon, and Y.-E.
  Sung, Nanorod-based dye-sensitized solar cells with improved charge
  collection efficiency, \emph{Adv. Mater.} {\bf 20}, \penalty0 54--58,
  (2008).

\bibitem{Adachi}
M.~Adachi, Y.~Murata, J.~Takao, J.~Jiu, M.~Sakamoto, and F.~Wang, Highly
  efficient dye-sensitized solar cells with a titania thin-film electrode
  composed of a network structure of single-crystal-like {TiO$_2$} nanowires
  made by the ``oriented attachment'' mechanism, \emph{J. Am. Chem. Soc.} {\bf
  126}, \penalty0 14943–--14949  (2004).

\bibitem{GandugliaPirovano2007219}
M.~V. Ganduglia-Pirovano, A.~Hofmann, and J.~Sauer, Oxygen vacancies in
  transition metal and rare earth oxides: Current state of understanding and
  remaining challenges, \emph{Surf. Sci. Rep.} {\bf 62}, \penalty0 219--270,
  (2007).

\bibitem{gatti2007}
M.~Gatti, F.~Bruneval, V.~Olevano, and L.~Reining, {Understanding correlations
  in vanadium dioxide from first principles}, \emph{{Phys. Rev. Lett.}} {\bf
  {99}}, \penalty0 266402--1--4  ({2007}).

\bibitem{kruger_prl}
P.~Kr\"{u}ger, S.~Bourgeois, B.~Domenichini, H.~Magnan, D.~Chandesris, P.~L.
  F\`{e}vre, A.~M. Flank, J.~Jupille, L.~Floreano, A.~Cossaro, A.~Verdini, and
  A.~Morgante, Defect states at the {TiO$_2$}(110) surface probed by resonant
  photoelectron diffraction, \emph{Phys. Rev. Lett.} {\bf 100}, \penalty0
  055501--1--4  (2008).

\bibitem{DiValentin}
C.~{Di Valentin}, G.~Pacchioni, and A.~Selloni, Electronic structure of defect
  states in hydroxylated and reduced rutile {TiO$_{2}$}(110) surfaces,
  \emph{Phys. Rev. Lett.} {\bf 97}, \penalty0 166803--1--4  (2006).

\bibitem{minato_jcp}
T.~Minato, Y.~Sainoo, Y.~Kim, H.~S. Kato, K.~ichi Aika, M.~Kawai, J.~Zhao,
  H.~Petek, T.~Huang, W.~He, B.~Wang, Z.~Wang, Y.~Zhao, J.~Yang, and J.~G. Hou,
  The electronic structure of oxygen atom vacancy and hydroxyl impurity defects
  on titanium dioxide (110) surface, \emph{J. Chem. Phys.} {\bf 130}, \penalty0
  124502--1--11  (2009).

\bibitem{finazzi_jcp}
E.~Finazzi, C.~D. Valentin, G.~Pacchioni, and A.~Selloni, Excess electron
  states in reduced bulk anatase {TiO$_2$}: Comparison of standard {GGA, GGA +
  U}, and hybrid {DFT} calculations, \emph{J. Chem. Phys.} {\bf 129}, \penalty0
  154113--1--9  (2008).

\bibitem{pacchioni_alone}
G.~Pacchioni, Modeling doped and defective oxides in catalysis with density
  functional theory methods: Room for improvements, \emph{J. Chem. Phys.} {\bf
  128}, \penalty0 182505--1--10  (2008).

\bibitem{Khomenko}
V.~M. Khomenko, K.~Langer, H.~Rager, and A.~Fett, Electronic absorption by
  {Ti$^{3+}$} ions and electron delocalization in synthetic blue rutile,
  \emph{Phys. Chem. Minerals} {\bf 25}, \penalty0 338--346  (1998).

\bibitem{StefanWendt06272008}
S.~Wendt, P.~T. Sprunger, E.~Lira, G.~K.~H. Madsen, Z.~Li, J.~O. Hansen,
  J.~Matthiesen, A.~Blekinge-Rasmussen, E.~Laegsgaard, B.~Hammer, and
  F.~Besenbacher, {The Role of Interstitial Sites in the {Ti}3$d$ Defect State
  in the Band Gap of Titania}, \emph{Science} {\bf 320}, \penalty0 1755--1759,
   (2008).

\bibitem{GW}
L.~Hedin, New method for calculating the one-particle {Green's} function with
  application to the electron-gas problem, \emph{Phys. Rev.} {\bf 139},
  \penalty0 A796--A823  (1965).

\bibitem{Dreizler}
R.~M. Dreizler and E.~K.~U. Gross, \emph{Density Functional Theory} (Springer
  Verlag, Heidelberg, 1990).

\bibitem{booksQP}
L.~Fetter and J.~D. Walecka, \emph{Quantum theory of Many Body Systems}
  (McGraw-Hill, New York, 1981).

\bibitem{Hohenberg}
P.~Hohenberg and W.~Kohn, Inhomogeneous electron gas, \emph{Phys. Rev.} {\bf
  136}, \penalty0 B864--B871  (1964).

\bibitem{Kohn}
W.~Kohn and L.~J. Sham, Self-consistent equations including exchange and
  correlation effects, \emph{Phys. Rev.} {\bf 140}, \penalty0 A1133--A1138,
  (1965).

\bibitem{LDA}
D.~M. Ceperley and B.~J. Alder, Ground state of the electron gas by a
  stochastic method, \emph{Phys. Rev. Lett.} {\bf 45}, \penalty0 566--569,
  (1980).

\bibitem{LDA2}
J.~P. Perdew and A.~Zunger, Self-interaction correction to density-functional
  approximations for many-electron systems, \emph{Phys. Rev. B} {\bf 23},
  \penalty0 5048--5079  (1981).

\bibitem{PBE}
J.~P. Perdew, K.~Burke, and M.~Ernzerhof, Generalized gradient approximation
  made simple, \emph{Phys. Rev. Lett.} {\bf 77}, \penalty0 3865--3868  (1996).

\bibitem{PBE0}
C.~Adamo and V.~Barone, Toward reliable density functional methods without
  adjustable parameters: The pbe0 model, \emph{J. Chem. Phys.} {\bf 110},
  \penalty0 6158--6170  (1999).

\bibitem{B3LYP}
A.~D. Becke, Density-functional thermochemistry. iii. the role of exact
  exchange, \emph{J. Chem. Phys.} {\bf 98}, \penalty0 5648--5652  (1993).

\bibitem{PerdewLevy}
J.~P. Perdew and M.~Levy, Physical content of the exact {Kohn-Sham} orbital
  energies: Band gaps and derivative discontinuities, \emph{Phys. Rev. Lett.}
  {\bf 51}, \penalty0 1884--1887  (1983).

\bibitem{booksQP2}
R.~D. Mattuck, \emph{A Guide to Feynman Diagrams in the Many-Body Problem}
  (McGraw-Hill, New York, 1976).

\bibitem{HedinLud}
L.~Hedin and S.~Lundqvist.
\newblock ``Effects of Electron-Electron and Electron-Phonon Interactions on the One-Electron States of Solids,''
\newblock in \emph{Solid State Physics}, Vol.~{\bf{23}}, edited by H.~Ehrenreich, F.~Seitz, and D.~Turnbull (Academic Press, New York, 1969) pp. 1--181. 

\bibitem{strinati}
G.~Strinati, Application of the {Green's} functions method to the study of the
  optical properties of semiconductors, \emph{Rivista Nuovo Cimento} {\bf 11},
  \penalty0 1--86  (1988).

\bibitem{Rubio}
P.~M. Echenique, J.~M. Pitarke, E.~V. Chulkov, and A.~Rubio, Theory of
  inelastic lifetimes of low-energy electrons in metals, \emph{Chem. Phys.}
  {\bf 251}, \penalty0 1--35  (2000).

\bibitem{hyblou}
M.~S. Hybertsen and S.~G. Louie, Electron correlation in semiconductors and
  insulators: Band gaps and quasiparticle energies, \emph{Phys. Rev. B} {\bf
  34}, \penalty0 5390--5413  (1986).

\bibitem{godshsh}
R.~W. Godby, M.~Schl\"uter, and L.~J. Sham, Self-energy operators and
  exchange-correlation potentials in semiconductors, \emph{Phys. Rev. B} {\bf
  37}, \penalty0 10159--10175  (1988).

\bibitem{AdlerWiser}
S.~L. Adler, Quantum theory of the dielectric constant in real solids,
  \emph{Phys. Rev.} {\bf 126}, \penalty0 413--420  (1962).

\bibitem{Lind}
J.~Lindhard, On the properties of a gas of charged particles, \emph{Mat. Fys.
  Medd. K. Dan. Vidensk. Selsk.} {\bf 28}, \penalty0 1--57  (1954).

\bibitem{review}
G.~Onida, L.~Reining, and A.~Rubio, Electronic excitations: density-functional
  versus many-body {Green's}-function approaches, \emph{Rev. Mod. Phys.} {\bf
  74}, \penalty0 601--659  (2002).

\bibitem{Hardman1994}
P.~J. Hardmand, G.~N. Raikar, C.~A. Muryn, G.~{van der Laan}, P.~L. Wincott,
  G.~Thornton, D.~W. Bullett, and P.~A. D. M.~A. Dale, Valence-band structure
  of {TiO$_{{2}}$} along the {$\Gamma$-$\Delta$-$X$ and $\Gamma$-$\Sigma$-$M$}
  directions, \emph{Phys. Rev. B} {\bf 49}, \penalty0 7170--7177  (1994).

\bibitem{Eriksen1987}
S.~Eriksen and R.~G. Egdell, Electronic excitations at oxygen deficient
  {TiO$_2$}(110) surfaces: A study by {EELS}, \emph{Surf. Sci.} {\bf 180},
  \penalty0 263--278  (1987).

\bibitem{Heise1992599}
R.~Heise, R.~Courths, and S.~Witzel, Valence band densities-of-states of
  {TiO$_2$}(110) from resonant photoemission and photoelectron diffraction,
  \emph{Solid State Commun.} {\bf 84}, \penalty0 599--602  (1992).

\bibitem{Egdell1986835}
R.~G. Egdell, S.~Eriksen, and W.~R. Flavell, Oxygen deficient {SnO$_2$} (110)
  and {TiO$_2$} (110): A comparative study by photoemission, \emph{Solid State
  Commun.} {\bf 60}, \penalty0 835--838  (1986).

\bibitem{Kurtz1989178}
R.~L. Kurtz, R.~Stock-Bauer, T.~E. Madey, E.~Rom{\'{a}}n, and J.~L.~D. Segovia,
  Synchrotron radiation studies of {H$_2$O} adsorption on {TiO$_2$}(110),
  \emph{Surf. Sci.} {\bf 218}, \penalty0 178--200  (1989).

\bibitem{TiO21953}
R.~G. Breckenridge and W.~R. Hosler, Electrical properties of titanium dioxide
  semiconductors, \emph{Phys. Rev.} {\bf 91}, \penalty0 793--802  (1953).

\bibitem{PhysRevLett.17.857}
F.~Arntz and Y.~Yacoby, Electroabsorption in rutile {TiO$_2$}, \emph{Phys. Rev.
  Lett.} {\bf 17}, \penalty0 857--860  (1966).

\bibitem{PhysRev.157.700}
A.~Frova, P.~J. Boddy, and Y.~S. Chen, Electromodulation of the optical
  constants of rutile in the {UV}, \emph{Phys. Rev.} {\bf 157}, \penalty0
  700--708  (1967).

\bibitem{PhysRev.184.979}
A.~K. Ghosh, F.~G. Wakim, and R.~R. Addiss, Photoelectronic processes in
  rutile, \emph{Phys. Rev.} {\bf 184}, \penalty0 979--988  (1969).

\bibitem{PhysRevB.51.6842}
A.~Amtout and R.~Leonelli, Optical properties of rutile near its fundamental
  band gap, \emph{Phys. Rev. B} {\bf 51}, \penalty0 6842--6851  (1995).

\bibitem{PhysRevB.7.2333}
S.~H\"ufner and G.~K. Wertheim, Core-electron splittings and hyperfine fields
  in transition-metal compounds, \emph{Phys. Rev. B} {\bf 7}, \penalty0
  2333--2336  (1973).

\bibitem{Fischer}
D.~W. Fischer, X-ray band spectra and molecular-orbital structure of rutile
  {TiO$_2$}, \emph{Phys. Rev. B} {\bf 5}, \penalty0 4219--4226  (1972).

\bibitem{PhysRevB.48.2074}
F.~M.~F. de~Groot, J.~Faber, J.~J.~M. Michiels, M.~T. Czy\ifmmode~\dot{z}\else
  \.{z}\fi{}yk, M.~Abbate, and J.~C. Fuggle, Oxygen 1$s$ x-ray absorption of
  tetravalent titanium oxides: A comparison with single-particle calculations,
  \emph{Phys. Rev. B} {\bf 48}, \penalty0 2074--2080  (1993).

\bibitem{Grunes}
L.~A. Grunes, Study of the $k$ edges of $3d$ transition metals in pure and
  oxide form by x-ray-absorption spectroscopy, \emph{Phys. Rev. B} {\bf 27},
  \penalty0 2111--2131  (1983).

\bibitem{PhysRevB.41.12366}
G.~van~der Laan, Polaronic satellites in x-ray-absorption spectra, \emph{Phys.
  Rev. B} {\bf 41}, \penalty0 12366--12368  (1990).

\bibitem{PhysRevB.40.5715}
F.~M.~F. de~Groot, M.~Grioni, J.~C. Fuggle, J.~Ghijsen, G.~A. Sawatzky, and
  H.~Petersen, Oxygen 1s x-ray-absorption edges of transition-metal oxides,
  \emph{Phys. Rev. B} {\bf 40}, \penalty0 5715--5723  (1989).

\bibitem{PhysRevB.46.2024}
K.~Watanabe, K.~Inoue, and F.~Minami, Resonant phenomena of
  hyper-raman-scattering of optic phonons in a {TiO$_2$} crystal, \emph{Phys.
  Rev. B} {\bf 46}, \penalty0 2024--2033  (1992).

\bibitem{PhysRev.87.876}
D.~C. Cronemeyer, Electrical and optical properties of rutile single crystals,
  \emph{Phys. Rev.} {\bf 87}, \penalty0 876--886  (1952).

\bibitem{Kavan}
L.~Kavan, M.~Gr{\"{a}}tzel, S.~E. Gilbert, C.~Klemenz, and H.~J. Scheel,
  Electrochemical and photoelectrochemical investigation of single-crystal
  anatase, \emph{J. Am. Chem. Soc.} {\bf 118}, \penalty0 6716--6723  (1996).

\bibitem{fleming:033707}
L.~Fleming, C.~C. Fulton, G.~Lucovsky, J.~E. Rowe, M.~D. Ulrich, and
  J.~L\"{u}ning, Local bonding analysis of the valence and conduction band
  features of {TiO$_2$}, \emph{J. Appl. Phys.} {\bf 102}, \penalty0
  033707--1--7  (2007).

\bibitem{sanjines:2945}
R.~Sanjin\'{e}s, H.~Tang, H.~Berger, F.~Gozzo, G.~Margaritondo, and
  F.~L\'{e}vy, Electronic structure of anatase {TiO$_2$} oxide, \emph{J. Appl.
  Phys.} {\bf 75}, \penalty0 2945--2951  (1994).

\bibitem{tang:2042}
H.~Tang, K.~Prasad, R.~Sanjin\`{e}s, P.~E. Schmid, and F.~L\'{e}vy, Electrical
  and optical properties of {TiO$_2$} anatase thin films, \emph{J. Appl. Phys.}
  {\bf 75}, \penalty0 2042--2047  (1994).

\bibitem{Cardona}
M.~Cardona and G.~Harbeke, Optical properties and band structure of
  wurtzite-type crystals and rutile, \emph{Phys. Rev.} {\bf 137}, \penalty0
  A1467--A1476  (1965).

\bibitem{Pascual}
J.~Pascual, J.~Camassel, and H.~Mathieu, Fine structure in the intrinsic
  absorption edge of {TiO$_2$}, \emph{Phys. Rev. B} {\bf 18}, \penalty0
  5606--5614  (1978).

\bibitem{Hosaka199675}
N.~Hosaka, T.~Sekiya, M.~Fujisawa, C.~Satoko, and S.~Kurita, Uv reflection
  spectra of anatase {TiO$_2$}, \emph{J. Electron Spec. Rel. Phen.} {\bf 78},
  \penalty0 75--78  (1996).

\bibitem{HosakaJ}
N.~Hosaka, T.~Sekiya, C.~Satoko, and S.~Kurita, Optical properties of
  single-crystal anatase {TiO$_2$}, \emph{J. Phys. Soc. Jpn.} {\bf 66},
  \penalty0 877--880  (1997).

\bibitem{Tang}
H.~Tang, F.~L{\'{e}}vy, H.~Berger, and P.~E. Schmid, Urbach tail of anatase
  {TiO$_2$}, \emph{Phys. Rev. B} {\bf 52}, \penalty0 7771--7774  (1995).

\bibitem{schilfgaarde:226402}
M.~van Schilfgaarde, T.~Kotani, and S.~Faleev, Quasiparticle self-consistent
  {GW} theory, \emph{Phys. Rev. Lett.} {\bf 96}, \penalty0 226402--1--4,
  (2006).

\bibitem{Labat1}
F.~Labat, P.~Baranek, C.~Doman, C.~Minot, and C.~Adamo, Density functional
  theory analysis of the structural and electronic properties of {TiO}$_{{2}}$
  rutile and anatase polytypes: Performances of different exchange-correlation
  functionals, \emph{J. Chem. Phys.} {\bf 126}, \penalty0 154703--1--12,
  (2007).

\bibitem{Muscat}
J.~Muscat, A.~Wander, and N.~M. Harrison, On the prediction of band gaps from
  hybrid functional theory, \emph{Chem. Phys. Lett.} {\bf 342}, \penalty0
  397--401  (2001).

\bibitem{Diebold200353}
U.~Diebold, The surface science of titanium dioxide, \emph{Surf. Sci. Rep.}
  {\bf 48}, \penalty0 53--229  (2003).

\bibitem{Chiodo}
L.~Chiodo, J.~M. Garc{\'{\i}}a-Lastra, A.~Incomino, H.~Petek, S.~Ossicini, and
  A.~Rubio.
\newblock \emph{Ab initio} electronic and optical description of {TiO$_2$}
  crystalline phases: Towards photocatalysis and photovoltaic applications.
\newblock (unpublished, 2009).

\bibitem{muscat2002}
J.~Muscat, V.~Swamy, and N.~Harrison, {First-principles calculations of the
  phase stability of TiO$_2$}, \emph{Phys. Rev. B} {\bf {65}}, \penalty0
  {224112--1--15}  ({2002}).

\bibitem{GW_TiO2_1}
M.~Oshikiri, M.~Boero, J.~Ye, F.~Aryasetiawan, and G.~Kido, The electronic
  structures of the thin films of {InVO}$_{{4}}$ and {TiO}$_{{2}}$ by first
  principles calculations, \emph{Thin Solid Films} {\bf 445}, \penalty0
  168--174  (2003).

\bibitem{PhysRevB.61.7459}
R.~Asahi, Y.~Taga, W.~Mannstadt, and A.~J. Freeman, Electronic and optical
  properties of anatase {TiO$_2$}, \emph{Phys. Rev. B} {\bf 61}, \penalty0
  7459--7465  (2000).

\bibitem{PhysRevB.64.184113}
M.~Calatayud, P.~Mori-S\'anchez, A.~Beltr\'an, A.~Mart\'in~Pend\'as,
  E.~Francisco, J.~Andr\'es, and J.~M. Recio, Quantum-mechanical analysis of
  the equation of state of anatase {TiO$_2$}, \emph{Phys. Rev. B} {\bf 64},
  \penalty0 184113--1--9  (2001).

\bibitem{GW_TiO2_anatase}
L.~Thulin and J.~Guerra, Calculations of strain-modified anatase {TiO}$_{{2}}$
  band structures, \emph{Phy. Rev. B} {\bf 77}, \penalty0 195112--1--5,
  (2008).

\bibitem{lawler:205108}
H.~M. Lawler, J.~J. Rehr, F.~Vila, S.~D. Dalosto, E.~L. Shirley, and Z.~H.
  Levine, Optical to {UV} spectra and birefringence of {SiO$_2$} and {TiO$_2$}:
  First-principles calculations with excitonic effects, \emph{Phys. Rev. B}
  {\bf 78}, \penalty0 205108--1--8  (2008).

\bibitem{BandGapErrors}
Z.-L. Cai, K.~Sendt, and J.~R. Reimers, Failure of density-functional theory
  and time-dependent density-functional theory for large extended $\pi$
  systems, \emph{J. Chem. Phys} {\bf 117}, \penalty0 5543--5549  (2002).

\bibitem{NewXCFuncs}
U.~Salzner, J.~B. Lagowski, P.~G. Pickup, and R.~A. Poirier, Design of low band
  gap polymers employing density functional theory -- hybrid functionals
  ameliorate band gap problem, \emph{J. Comput. Chem.} {\bf 18}, \penalty0
  1943--1953  (1997).

\bibitem{DeAngelis}
F.~{De Angelis}, A.~Tilocca, and A.~Selloni, Time-dependent dft study of
  {[Fe(CN)$_6$]$^{4-}$} sensitization of {TiO$_2$} nanoparticles, \emph{J. Am.
  Chem. Soc.} {\bf 126}, \penalty0 15024--15025  (2004).

\bibitem{RPBE}
B.~Hammer, L.~B. Hansen, and J.~K. N{\/{o}}rskov, Improved adsorption
  energetics with density functional theory using revised
  {Perdew}-{Burke}-{Ernzerhof} functionals, \emph{Phys. Rev. B} {\bf 59},
  \penalty0 7413--7421  (1999).

\bibitem{Gratzel2003}
M.~Gratzel, {Dye-sensitized solar cells}, \emph{J. Photochem. Photobiol.
  C--Photochem. Rev.} {\bf {4}}, \penalty0 {145--153}  ({2003}).

\bibitem{tddftbook}
M.~A.~L. Marques, C.~Ullrich, F.~Nogueira, A.~Rubio, K.~Burke, and E.~Gross,
  Eds., \emph{Time-Dependent Density Functional Theory}, Vol.~{\bf{706}}, (Springer,
  Berlin, 2006).

\bibitem{deangelis2004}
F.~De~Angelis, A.~Tilocca, and A.~Selloni, {Time-dependent DFT study of
  {[}Fe(CN)(6)](4-) sensitization of TiO$_2$ nanoparticles}, \emph{{J. Am.
  Chem. Soc.}} {\bf {126}}, \penalty0 {15024--15025}  ({2004}).

\bibitem{deangelis2005}
F.~De~Angelis, S.~Fantacci, A.~Selloni, and M.~Nazeeruddin, {Time dependent
  density functional theory study of the absorption spectrum of the
  {[}Ru(4,4'-COO$^-$-2,2'-bpy)$_2$(X)$_2$]$^{4-}$ (X = NCS, Cl) dyes in water
  solution}, \emph{{Chem. Phys. Lett.}} {\bf {415}}, \penalty0 {115--120},
  ({2005}).

\bibitem{deangelis2005-2}
M.~Nazeeruddin, F.~De~Angelis, S.~Fantacci, A.~Selloni, G.~Viscardi, P.~Liska,
  S.~Ito, B.~Takeru, and M.~Gratzel, {Combined experimental and DFT-TDDFT
  computational study of photoelectrochemical cell ruthenium sensitizers},
  \emph{{J. Am. Chem. Soc.}} {\bf {127}}, \penalty0 {16835--16847}  ({2005}).

\bibitem{deangelis2007}
F.~De~Angelis, S.~Fantacci, A.~Selloni, M.~K. Nazeeruddin, and M.~Gratzel,
  {Time-dependent density functional theory investigations on the excited
  states of {Ru(II)}-dye-sensitized {TiO$_2$} nanoparticles: The role of
  sensitizer protonation}, \emph{{J. Am. Chem. Soc.}} {\bf {129}}, \penalty0
  {14156--14157}  ({2007}).

\bibitem{deangelis2008}
F.~De~Angelis, S.~Fantacci, and A.~Selloni, {Alignment of the dye's molecular
  levels with the {TiO$_2$} band edges in dye-sensitized solar cells: a
  {DFT-TDDFT} study}, \emph{{Nanotechnology}} {\bf {19}}, \penalty0
  {424002--1--7}  ({2008}).

\bibitem{prezhdo2005}
W.~R. Duncan, W.~M. Stier, and O.~V. Prezhdo, \emph{Ab initio} nonadiabatic
  molecular dynamics of the ultrafast electron injection across the
  alizarin-−{TiO$_2$} interface, \emph{{J. Am. Chem. Soc.}} {\bf {127}},
  \penalty0 {7941--7951}  ({2005}).

\bibitem{prezhdo2007}
W.~R. Duncan and O.~V. Prezhdo, {Theoretical studies of photoinduced electron
  transfer in dye-sensitized TiO$_2$}, \emph{{Ann. Rev. Phys. Chem.}} {\bf
  {58}}, \penalty0 {143--184}  ({2007}).

\bibitem{prezhdo2007-2}
W.~R. Duncan, C.~F. Craig, and O.~V. Prezhdo, {Time-domain {\emph{ab initio}}
  study of charge relaxation and recombination in dye-sensitized TiO$_2$},
  \emph{{J. Am. Chem. Soc.}} {\bf {129}}, \penalty0 {8528--8543}  ({2007}).

\bibitem{prezhdo2008}
O.~V. Prezhdo, W.~R. Duncan, and V.~V. Prezhdo, {Dynamics of the photoexcited
  electron at the chromophore-semiconductor interface}, \emph{{Acc. Chem.
  Res.}} {\bf {41}}, \penalty0 {339--348}  ({2008}).

\bibitem{prezhdo2008-2}
W.~R. Duncan and O.~V. Prezhdo, {Temperature independence of the photoinduced
  electron injection in dye-sensitized {TiO$_2$} rationalized by \emph{ab
  initio} time-domain density functional theory}, \emph{{J. Am. Chem. Soc.}}
  {\bf {130}}, \penalty0 {9756--9762}  ({2008}).

\end{thebibliography}

\end{document}